\documentclass[11pt]{article}
\pdfoutput=1  % togliere il commento se le figure sono pdf, lasciarlo se sono eps
\usepackage{graphicx}

\usepackage{appendix}

\usepackage{amssymb}
\usepackage{amsmath}
\usepackage{amsfonts}
\usepackage{upgreek}
\usepackage{latexsym}
\usepackage{slashed,upgreek,amscd,cancel,tensor}
\usepackage{adjustbox}
\usepackage{epsfig}
\usepackage{cite}
\usepackage{hyperref}
\usepackage{amsfonts}
\usepackage{booktabs}
\usepackage{siunitx}
\usepackage{caption}
\usepackage{float}
\usepackage{color}

\numberwithin{equation}{section}% numera le equazioni seconde le sezioni , e.g. 1.15 invece che consecutivamente; anche le appendici, eq. (A.1) etc. Richiede amsmath

\definecolor{MyBlue}{rgb}{0.15,0.15,0.70}

\hypersetup{
colorlinks=true,
citecolor=MyBlue,
linkcolor=MyBlue,
urlcolor=MyBlue
}

\input epsf
\usepackage[top=2cm,bottom=3cm,left=2.6cm,right=2.6cm,asymmetric]{geometry}

\newcommand{\be}{\begin{equation}}
\newcommand{\ee}{\end{equation}}
\newcommand{\beq}{\begin{equation}}
\newcommand{\eeq}{\end{equation}}
\newcommand{\bea}{\begin{eqnarray}}
\newcommand{\eea}{\end{eqnarray}}

\usepackage[stable]{footmisc}

\def\dkmu2{\delta K_{\mu \nu}\delta K^{\mu \nu}}
\def\pmu2{  \phi_{\mu \nu}\phi^{\mu \nu}}

\newcommand\cL{{\mathcal{L}}}
\newcommand\cS{{\mathcal{S}}}
\newcommand\cT{{\mathcal{T}}}
\newcommand\cJ{{\mathcal{J}}}
\newcommand\cK{{\mathcal{K}}}
\newcommand\cR{{\mathcal{R}}}
\newcommand\cO{{\mathcal{O}}}
\newcommand\cC{{\mathcal{C}}}

\renewcommand\[{\left[}

\newcommand\ees{\end{eqnarray}}
\newcommand\bees{\begin{eqnarray}}

\begin{document}

\begin{center}
\LARGE{\bf Hairy black holes in DHOST theories: \\ Exploring disformal transformation as a solution generating method}
\\[1cm] 

\large{Jibril Ben Achour$^{\rm a}$, Hongguang Liu$^{\rm b,c,d}$, Shinji Mukohyama$^{{\rm e},{\rm a}}$}
\\[0.5cm]

\small{
\textit{$^{\rm a}$ Center for Gravitational Physics, Yukawa Institute for Theoretical Physics \\ [0.05cm]
Kyoto University, Kyoto, Japan}}

\vspace{.2cm}

\small{
\textit{$^{\rm b}$ Aix Marseille Univ, Universit\'e de Toulon, CNRS, CPT,
\\[0.05cm]
 13000 Marseille, France}}

\vspace{.2cm}

\small{
\textit{$^{\rm c}$ Laboratoire de Math\'ematiques et Physique Th\'eorique, \\ [0.05cm]
Universit\'e Fran\c cois Rabelais, 
\\[0.05cm]
Parc de Grandmont, 37200 Tours, France}}

\vspace{.2cm}

\small{
\textit{$^{\rm d}$ Center for Quantum Computing, Pengcheng Laboratory, \\ [0.05cm] 
Shenzhen 518066, China}}

\vspace{.2cm}

\small{
\textit{$^{\rm e}$ Kavli Institute for the Physics and Mathematics of the Universe (WPI), \\ [0.05cm]
The University of Tokyo Institutes for Advanced Study, The University of Tokyo
\\[0.05cm]
Chiba 277-8583, Japan}}

 %CEA, IPhT, 91191 Gif-sur-Yvette c\'edex, France \\ [0.05cm]
 %\\
%CNRS,  URA-2306, 91191 Gif-sur-Yvette c\'edex, France}}
\end{center}

\vspace{.2cm}

\begin{abstract}
Solutions-generating methods based on field redefinitions, such as conformal mapping, play an important role in investigating exact solutions in modified gravity. In this work, we explore the possibility to use disformal field redefinitions to investigate new regions of the solution space of DHOST theories, and present new hairy black holes solutions beyond the stealth sector. The crucial ingredient is to find suitable seed solutions to generate new exact ones for DHOST theories. We first consider a seed solution of the Einstein-Scalar system describing a naked singularity. Under suitable assumptions, we derive a no-go result showing that no black hole solution can be constructed from such a seed GR solution. Then, taking into account the stability of each degeneracy classes of quadratic DHOST theories under a general disformal mapping, we consider two kinds of known black hole solutions as seed configurations: the Schwarzschild stealth solution and the non-stealth Reissner-Nordstrom like solution. Restricting our considerations to invertible disformal transformations, we show that building new hairy black hole solutions from the stealth solution associated to a constant kinetic term is also quite constrained. We obtain new solutions which either are stealth or describe asymptotically locally flat black holes with a deficit solid angle. However, starting from the non-stealth seed solution associated to a non constant kinetic term as well as a time-dependent scalar profile, we show that a disformal transformation can introduce rather general modifications of the exterior geometry. Finally, we consider the construction of minimally modified hairy black hole solutions using a small disformal transformation. Further applications of this solution generating method should allow to provide new hairy rotating black hole solutions beyond the stealth sector, as well as minimally modified GR solutions useful for phenomenological investigations of compact objects beyond GR.

\end{abstract}
\qquad \;\;\;YITP-19-87, IPMU-19-0130
\vspace{.2cm}

\newpage

\tableofcontents

\newpage

\section{Introduction}

General Relativity (GR) is up to now our best effective theory of the gravitational field. Yet, in order to fit the observations at cosmological and galactic scales without modifying GR, one has to introduce new unknown sources of gravity dubbed the dark sector, leading finally to our current $\Lambda\text{CDM}$ model of cosmology\footnote{Nevertheless, it is fair to say that the status of the $\Lambda\text{CDM}$ is now under debate due to the $H_{0}$ tension. See \cite{Bernal:2016gxb} for a review.} \cite{Bull:2015stt, Huterer:2017buf, Ezquiaga:2018btd}. Modified gravity theories aiming at providing instead infrared modifications of gravity, while restoring GR on local scales, have been extensively developed since then as an alternative path to explain the dark sector. Among them, scalar-tensor theories are by far the most studied candidates. A common consistency criterion for such modified gravity theories is not to introduce Ostrogradsky ghosts, which could make the theory unstable. The most general scalar-tensor theory admitting second order field equations and which avoids the existence of an Ostrogradsky ghost was constructed by Horndeski in \cite{Horndeski:1974wa} and rediscovered in \cite{Deffayet:2009wt, Deffayet:2009mn, Deffayet:2011gz} afterwards. While having no more than second order derivatives in field equations has remained for some time a criterion for avoiding the Ostrogradsky ghost, healthy theories beyond Horndeski with higher order field equations, known as the GLPV theories, were discovered in \cite{Gleyzes:2014dya, Gleyzes:2014qga}, suggesting the existence of a broader class of theories beyond Horndeski and thus of a more general criterion to build healthy theories. 

This idea was concretely realized in \cite{Langlois:2015cwa} and developed in a series of subsequent works \cite{Langlois:2015skt, Achour:2016rkg, BenAchour:2016fzp, Motohashi:2016ftl, Crisostomi:2016czh, Crisostomi:2017aim}. It turned out that the relevant criterion to avoid the Ostrogradsky ghost is to work with degenerate Lagrangians. The degeneracy property reflects the existence of an additional constraint at the phase space level, which can be used to remove the unwanted ghost \cite{Langlois:2015skt}. Degeneracy conditions therefore provide an efficient way to build consistent extensions of GR, even beyond the scalar-tensor framework \cite{Heisenberg:2018vsk}. These degenerate higher order scalar-tensor (DHOST) theories encompass a large set of relevant scalar-tensor theories developed so far, and thus provide a landscape of healthy scalar-tensor theories which can be efficiently confronted with observations. Recently, the landscape of DHOST theories has been severely restricted by the recent detection of the joint event GW170817-GR170817 \cite{Langlois:2017dyl}, and more recently by excluding graviton decay into scalar field perturbations \cite{Creminelli:2018xsv}.  These constraints led to a restricted space of theories consistent with the observations, at least on cosmological scales\footnote{It is worth pointing that although the original motivation was to provide infrared modifications of GR, the energy scale at which DHOST theories can be used as an effective theory of gravity beyond GR is not fixed a priori. In principle, this landscape of healthy theories could be used to model deviation from GR in the UV as well. An example is given by mimetic gravity and its extension \cite{Chamseddine:2014vna}, which can be reformulated in the DHOST framework \cite{Langlois:2018jdg}. This theory and its extensions have been used to provide effective regular cosmological and black hole models \cite{Chamseddine:2016uef, Chamseddine:2016ktu, Yoshida:2018kwy, Brahma:2018dwx}, as well as covariant action for effective quantum cosmology and quantum black holes models inspired by loop quantum gravity \cite{Liu:2017puc, Bodendorfer:2017bjt, Bodendorfer:2018ptp, BenAchour:2017ivq, deCesare:2019pqj}.}. See \cite{Langlois:2018dxi, Langlois:2017mdk, Kobayashi:2019hrl} for recent reviews.

 Over the past five years, important efforts have been devoted in investigating the deviations from GR induced by DHOST theories on compact objects. Early works on the structure of stars were presented in \cite{Cisterna:2015yla, Cisterna:2016vdx, Babichev:2016jom, Sakstein:2016oel} and more recently in \cite{ Lehebel:2017fag, Kobayashi:2018xvr, Babichev:2018rfj, Lehebel:2018zga}.  As for exact black hole solutions, no-hair theorems were found for shift symmetric Horndeski \cite{Hui:2012qt} and later on for GLPV theories \cite{Babichev:2017guv}. These generalizations of the no-hair theorem helped to identify the conditions to relax in order to build hairy black hole solutions. See \cite{Babichev:2016rlq} for more details. Among such conditions, the introduction of a linearly time-dependent scalar profile has allowed the construction of a large class of hairy black hole solutions \cite{Babichev:2012re, Anabalon:2013oea, Babichev:2013cya, Charmousis:2014zaa, Kobayashi:2014eva, Babichev:2016kdt, Motohashi:2018wdq, Minamitsuji:2018vuw}. Yet, up to now, most of the exact solutions found so far correspond to stealth configurations, which consist of GR vacuum solutions dressed by a scalar field with a vanishing energy momentum tensor. A rather generic property of these solutions is to have a constant value of the scalar kinetic term $X = g^{\mu\nu}\partial_{\mu}\phi \partial_{\nu} \phi$.  The general conditions for the existence of such stealth solutions were systematically derived for quadratic DHOST theories in \cite{BenAchour:2018dap, Motohashi:2019sen}. See also \cite{Minamitsuji:2019shy}. More recently, a rotating stealth solution was derived in \cite{Charmousis:2019vnf}. Stability and strong coupling issues of these stealth solutions were investigated in \cite{Babichev:2017lmw, Babichev:2018uiw} and even more recently in \cite{Takahashi:2019oxz, deRham:2019gha, Charmousis:2019fre}. Black hole solutions that are approximately stealth and are free from the strong coupling problem were also found and studied in the context of ghost condensation in \cite{Mukohyama:2005rw,Mukohyama:2009rk,Mukohyama:2009um, ArkaniHamed:2003uy, ArkaniHamed:2003uz}, and can be extended to more general setups \cite{Motohashi:2019ymr}. Beyond the stealth sector, only few exact black hole solutions have been found. See \cite{Babichev:2017guv}, \cite{Babichev:2013cya} and \cite{Kobayashi:2014eva} for useful examples. Summarizing this discussion, all these fruitful exact solutions, being stealth or not, have been obtained using the generalization of the no hair theorem to identify simple conditions to violate in order to derive hairy black hole solutions. 
 In this work, we would like to explore yet another approach to construct new hairy black hole solutions for DHOST theories.
 
 Following previous solution generating method in scalar-tensor theories based on a conformal transformation \cite{Bekenstein:1974sf, Sneddon, Yazadjiev:2001bx, Faraoni:2016ozb, Faraoni:2017afs, Faraoni:2017ecj, Chauvineau:2018zjy}, in particular in Brans-Dicke theory, we explore the possibility to generate new exact black hole solutions of DHOST theories using a disformal transformation of known exact solutions. See \cite{Bekenstein:1992pj, Ezquiaga:2017ner, Zumalacarregui:2010wj, Zumalacarregui:2013pma} for details on disformal transformations. They key point is that each degeneracy class of quadratic DHOST theories are stable under a general disformal transformation. Therefore, knowing an exact solution in one class, it is straightforward to generate a new exact solution for another representative DHOST theory belonging to the same class. Moreover, since the Einstein-Scalar system is mapped to a subset of DHOST theories under a disformal transformation, it is also possible to generate a new exact solution for DHOST theories starting from a seed solution of the Einstein-Scalar system. 
 
 The crucial requirement to apply this strategy is the existence of a seed solution. In this work, we shall implement the disformal transformation on three different types of seed solutions:
 \begin{itemize}
 \item  (I) The prototypical solution of the Einstein-Scalar system describing a naked singularity. Due to the no hair theorem, such naked singularity solutions are the only available ones when restricting to a canonical scalar field. 
 \item (II) The standard Schwarzschild stealth solution of quadratic DHOST theories associated to a linearly time-dependent scalar profile and a constant kinetic term. 
 \item (III) A non-stealth solution of GLPV theories which is also associated to time-dependent scalar profile but with a non-constant kinetic term.
 \end{itemize}
 These three types of exact solutions are interesting in that they provide explicit examples of solutions which satisfy all or part of the following three crucial properties  i) the existence of a black hole horizon, ii) the existence of a time-dependent scalar profile and finally iii) the existence of a non-constant kinetic term . As we will show, this allows to test how the following properties of the seed solution affect the construction of exact black hole solutions. In the following, we shall show that dressing the asymptotically flat naked singularity of the Einstein-Scalar system with a black hole horizon is impossible under suitable assumptions, leading to a no-go result when starting with (I). Additionally, if one considers a seed solution of type (II) then the constancy of the kinetic term associated to the stealth solution implies that the two disformal potentials are also constant, leaving not much freedom to construct new black hole solution. Finally, we will show that the last example (III) allows for a much larger freedom in constructing modified black hole solutions, the main reason being the combination of a time-dependent scalar profile and a non-constant kinetic term. 

 In the end, our investigation provides a first implementation of this solution generating method in the context of DHOST, focusing on spherically symmetric and static geometries with a possibly time-dependent scalar profile. We point that the same idea was already implemented in the context of Vector-Tensor theories in \cite{Filippini:2017kov}. Starting from other seed solutions, further applications of this strategy shall allow for the construction of more interesting exact solutions. In particular, one could generate new rotating black hole solutions starting from the stealth Kerr solution found in \cite{Charmousis:2019vnf}, as well as more subtle geometries such as the Robinson-Trautman-like solution, which describes non-linear radiative gravitational-field. Hence, this work only provides the first step towards a wider exploration of this solution generating method.

This article is organized as follows. In Section-\ref{sec1}, we recall the disformal mapping of quadratic DHOST theories as well as the subsector which is consistent with the recent gravitational waves observations and which avoids the graviton decay problem. In Section-\ref{sec1.3}, we discuss the sub-sector of DHOST which does not satisfy the observational constraints but which can be mapped to the viable subsector by a general disformal transformation. In Section-\ref{sec3}, we discuss the disformal transformation of the Janis-Newman-Winicour solution of the Einstein-Scalar system which describes a naked singularity, and present the detailed proof of our no-go result.  Then, Section-\ref{sec4} is devoted to the disformal transformation of the seed solution within the DHOST framework. Finally, Section-\ref{sec6} is devoted to a summary of the paper and discussions. We conclude by a discussion of our results and the perspectives opened by our work.

\section{Disformal mapping of the quadratic DHOST action}

\label{sec1}

In this section, we review the disformal transformation of the quadratic DHOST action and recall the stability of each DHOST sub-class under the disformal transformation. More details on this aspect can be found in \cite{Achour:2016rkg}. After recalling the restriction in order to be consistent with the recent observational constraints, we derive the constraint on the coupling constants of a general quadratic DHOST theories to be mapped to the viable sub-sector consistent with the observation via a disformal transformation.

\subsection{Reviewing the general results}

Consider the quadratic DHOST action, given by
\begin{align}
\label{DHOST}
\tilde{\cS}[\phi,  \tilde{g}_{\mu\nu}]  = \int d^4x \sqrt{|\tilde{g}|} \left\{ \tilde{f} \tilde{R} + \tilde{\cL}_{0} + \sum_{I=1}^5  \tilde{\alpha}_{I}  \tilde{\cL}^{(\phi)}_{I} \right\}\,,
\end{align}
where the Lagrangian $\tilde{\cL}_{0}$ is given by
\be
\tilde{\cL}_{0} =  \tilde{P} + \tilde{Q} \; \tilde{g}^{\mu\nu} \tilde{\nabla}_{\mu} \tilde{\nabla}_{\nu}\phi
\ee
while the other scalar field Lagrangians are given by
\begin{align}
 \tilde{\cL}^{\phi}_1 & = \tilde{\nabla}^{\mu} \tilde{\nabla}^{\nu} \phi \tilde{\nabla}_{\mu} \tilde{\nabla}_{\nu}  \phi\,,\\
\tilde{\cL}^{\phi}_2 & = \left( \tilde{g}^{\mu\nu} \tilde{\nabla}_{\mu} \tilde{\nabla}_{\nu}\phi\right)^2\,, \\
\tilde{ \cL}^{\phi}_3 & = \tilde{g}^{\rho\sigma} \tilde{\nabla}_{\rho} \tilde{\nabla}_{\sigma} \phi \tilde{\nabla}_{\mu} \tilde{\nabla}_{\nu} \phi  \tilde{\nabla}^{\mu} \phi \tilde{\nabla}^{\nu} \phi\,, \\
 \tilde{ \cL}^{\phi}_4 & = \tilde{\nabla}^{\mu}\phi \tilde{\nabla}_{\mu} \tilde{\nabla}_{\nu}  \phi  \tilde{\nabla}^{\nu} \tilde{\nabla}^{\rho} \phi  \tilde{\nabla}_{\rho} \phi\,, \\
 \tilde{  \cL}^{\phi}_5 & = \left( \tilde{\nabla}^{\mu}\phi \tilde{\nabla}_{\mu} \tilde{\nabla}_{\nu} \phi \tilde{\nabla}^{\nu}\phi \right)^2\,.
\end{align}
Here, $\tilde{f}$, $\tilde{P}$, $\tilde{Q}$ and $\tilde{\alpha}_I$ ($I=1,\cdots,5$) are functions of $\phi$ and $\tilde{X} \equiv \tilde{g}^{\mu\nu}\partial_{\mu}\phi\partial_{\nu}\phi$, and $\tilde{\nabla}$ is the covariant derivative compatible with the metric $\tilde{g}_{\mu\nu}$. Degeneracy conditions for this DHOST lagrangian are given by three conditions which can be written as $D_0 = D_1 =D_2 =0$. For simplicity, we only report the first condition, the two other ones can be found in \cite{Langlois:2015cwa}. One has
\begin{align}
D_{0} = \left( \tilde{\alpha}_1 + \tilde{\alpha}_2 \right) \left( \tilde{X} \tilde{f} \left( 2 \tilde{\alpha}_1 + \tilde{X} \tilde{\alpha}_4 + 4\tilde{f}_{\tilde{X}}\right) - 2 \tilde{f}^2 - 8 \tilde{X}^2 \tilde{f}^2_{\tilde{X}} \right) =0\;,
\end{align}
which leads to the simplest degeneracy condition
\be
\label{deg1}
\tilde{\alpha}_1 + \tilde{\alpha}_2 =0
\ee
Starting from this condition, one can then classify the space of theories which are degenerate using the remaining conditions $D_1 =D_2 =0$. See \cite{Langlois:2015cwa} for details.

Let us now discuss the property of this Lagrangian under disformal transformations. The disformal transformation of this DHOST action has been discussed in detail in \cite{Achour:2016rkg}, and we shall only review the main results in this section. Noticed that the Lagrangian $\tilde{\cL}_{0}$ does not contribute to the degeneracy conditions and we will not discuss its transformation in the following. Let us consider a general disformal transformation which acts on the metric as
\be
\label{dm}
\tilde{g}_{\mu\nu} = A\left( \phi, X\right) g_{\mu\nu} + B\left( \phi, X\right) \phi_{\mu} \phi_{\nu}\,,
\ee
but leaves the scalar field invariant. See \cite{Bekenstein:1992pj, Ezquiaga:2017ner, Zumalacarregui:2010wj, Zumalacarregui:2013pma} for details on these fields redefinitions. Performing this mapping on the action (\ref{DHOST}), one obtains a new action of the form
\be
\label{LAG}
\cS[ \phi,  g_{\mu\nu}]  = \int d^4x \sqrt{| g|} \left\{ f  R + \sum_{I=1}^5  \alpha_{I}  \cL^{(\phi)}_{I} \right\}\,,
\ee
where the new functions $f(\phi, X)$ and $\alpha_{I}(\phi, X)$ are given in term of the old ones by
\begin{align}
\label{1}
f & = \cJ_{g} A^{-1} \tilde{f}\,, \nonumber \\
\alpha_1 & = - h + \cJ_g \cT_{11} \tilde{\alpha}_1\,, \nonumber \\
\alpha_2 & = h + \cJ_g \cT_{22} \tilde{\alpha}_2\,, \nonumber \\
 \alpha_3 & = 2h_{X} \nonumber  + \cJ_g \left[ \tilde{f} \gamma_3 - 2 \tilde{X}_X \tilde{f}_{\tilde{X}} \lambda_3 + \cT_{13} \tilde{\alpha}_1 + \cT_{23} \tilde{\alpha}_2 + \cT_{33} \tilde{\alpha_3 }\right]\,, \nonumber  \\
 \alpha_4 & = - 2h_{X}  + \cJ_g \left[ \tilde{f} \gamma_4 - 2 \tilde{X}_X \tilde{f}_{\tilde{X}} \lambda_4 +  \cT_{14} \tilde{\alpha}_1 + \cT_{44} \tilde{\alpha}_4 \right]\,,\nonumber \\
\alpha_5 & = \cJ_g \left[ \tilde{f} \gamma_5 - 2 \tilde{X}_X  \tilde{f}_{\tilde{X}} \lambda_5 + \cT_{15} \tilde{\alpha}_1 + \cT_{25} \tilde{\alpha}_2 + \cT_{35} \tilde{\alpha}_3 + \cT_{45} \tilde{\alpha}_4+ \cT_{55}  \tilde{\alpha}_5 \right]\,,
\end{align}
where the precise form of the functions $\gamma_I$, $\lambda_I$ and the coefficients $\cT_{IJ}$ are given in \cite{Achour:2016rkg}.
The functions $\cJ_g(\phi, X)$ and $h(\phi, X)$ are given by
\begin{align}
\cJ_g   = \frac{\sqrt{|\tilde{g}|}}{\sqrt{|g|}}  = A^{3/2} \sqrt{A + B X}\,, \qquad h   = - \frac{B}{A\left( A + B X\right)} \cJ_g \tilde{f}\,.
\end{align}
Finally, the kinetic term transforms as
\be
\label{Xtilde}
\tilde{X} = \frac{X}{A + B X}\,. 
\ee
A useful property is that each of the degeneracy class of DHOST theories, classified by their degeneracy conditions, is stable under this mapping. As a simple example, the transformation of the degeneracy condition (\ref{deg1}) reads
\be
\alpha_1 + \alpha_2 = \cJ_g \cT_{11} \left( \tilde{\alpha}_1 + \tilde{\alpha}_{2}\right) = 0
\ee
The transformations of the other degeneracy conditions can be found in \cite{Achour:2016rkg}. As a result, the disformal transformation of a quadratic DHOST theory leads to a new quadratic DHOST theory belonging to the same degeneracy class.
As such, each degeneracy class can be viewed as an equivalence class of DHOST theories under a disformal mapping. From the point of view of the solution-generating method, this stability property is useful in that it ensures that knowing an exact solution of a DHOST theory, one can generate by a disformal transformation a new exact solution of another DHOST theory which lives in the same degeneracy class than the initial one.

\subsection{Remark on the invertibility of the disformal transformation}

\label{invertibility}

In the following, we shall be interested in invertible mapping between exact solutions of the form
\be
\left( \tilde{g}_{\mu\nu} , \phi \right) \; \qquad \rightarrow \qquad \left( g_{\mu\nu} , \phi  \right)\,.
\ee
The invertibility of this mapping requires that the Jacobian determinant is non-zero, 
\begin{align}
\left|\frac{\partial \tilde{g}_{\mu\nu}}{\partial g_{\mu\nu}} \right| \neq 0\,. 
\end{align}
This is equivalent to the requirement that all eigenvalues of the Jacobian be non-vanishing, resulting in~\cite{Zumalacarregui:2013pma}
\begin{equation}
 A - X A_X - X^2 B_X \ne 0\,, \quad A \ne 0\,. \label{eqn:invertibility}
\end{equation}
The first condition can be rewritten as $\partial\tilde{X}/\partial X \ne 0$, where $\tilde{X}$ is considered as a function of $\phi$ and $X$ through the relation (\ref{Xtilde}).

If $\partial_{\mu}\phi$ is timelike before and after the transformation then there is a simple interpretation of the invertibility condition (\ref{eqn:invertibility})~\footnote{One can easily find a similar interpretation if $\partial_{\mu}\phi$ is spacelike before and after the transformation.}. In this case one can take the unitary gauge where $\phi=t$, and in this gauge the lapse functions for the two metrics $\tilde{g}_{\mu\nu}$ and $g_{\mu\nu}$ are $\tilde{N}=1/\sqrt{-\tilde{X}}$ and $N=1/\sqrt{-X}$, respectively. Therefore the first condition in (\ref{eqn:invertibility}) is nothing but the invertibility of the mapping between the two lapse functions. On the other hand, the mapping between the spatial metrics in the unitary gauge is a conformal transformation and the second condition in (\ref{eqn:invertibility}) tells that the conformal factor for this conformal transformation should be non-vanishing. 

Throughout the present paper we shall impose the invertibility condition (\ref{eqn:invertibility}) on the disformal transformation.

\subsection{Observational constraints}
\label{subsec:consraints}

Before presenting the solution generating method, let us briefly recall the subset of DHOST theories which is consistent with the recent observational constraints. 

For cosmological backgrounds with non-vanishing $X$, following \cite{Langlois:2017dyl}, the recent joined detection GW081708-GR081708 imposing that the speed of gravitational waves be the same as the speed of light, up to $10^{-15}$, at least on cosmological scale, leads to the following constraints on the coupling functions of the DHOST Lagrangian
\begin{align}
\alpha_1 = \alpha_2 =0\,,
\end{align}
as well as 
\begin{align}
\label{aa4}
\alpha_4 &= \frac{1}{8f} \left[ 48 f^2_X - 8 (f-X f_X) \alpha_3 - X^2 \alpha^2_3\right]\,,\\
\label{aa5}
\alpha_5 &= \frac{1}{2f} \left[ 4f_X + X \alpha_3\right] \alpha_3\,.
\end{align}
These conditions are derived as follows. First, requiring that $c_g =c$ imposes $\alpha_1 =0$. Then using the degeneracy conditions derived in \cite{Langlois:2015cwa}, one obtains $\alpha_2=0$ as well as (\ref{aa4}) and (\ref{aa5}), while $\alpha_3$ remains free. See \cite{Langlois:2017dyl} for details.

Moreover, following \cite{Creminelli:2018xsv}, the restriction imposing that gravitational waves do not decay into scalar field perturbations requires that $\alpha_3 =0$ which strengthens the constraints to
\be
\label{Gdecay}
\alpha_1= \alpha_2 = \alpha_3 = \alpha_5 =0 \;, \qquad \alpha_4 = \frac{6f^2_X}{f}\,. 
\ee

In the present case, generating new exact solutions to the DHOST field equations via a disformal mapping allows actually to work beyond this subsector. Indeed, in order for our new exact solutions to satisfy the constraint $c_g=c$, only the final target theory has to belong to the viable sub-sector of DHOST theories. However, not all DHOST theories can be mapped back to this sub-sector by a disformal transformation. In the next section, we derive the constraint on the initial DHOST theory such that its image under a disformal mapping satisfies $c_g=c$.

\subsection{From general DHOST to the viable sector via disformal mapping}

\label{sec1.3}

Consider now a disformal transformation of a general quadratic DHOST theory. We would like to derive the condition for a general DHOST theory to be mapped to the viable sub-sector presented above under a disformal transformation. In order to satisfy the constraints (\ref{Gdecay}) after disformal transformation, one obtains from (\ref{1}) that the coupling functions of the initial DHOST theory have to satisfy the following conditions. The first coupling function $\tilde{\alpha}_1$ reads
\be
\label{condd1}
\tilde{\alpha}_1 = \frac{h}{\cJ_g \cT_{11}} = - \frac{B \left( A + B X\right)}{A} \tilde{f}\,,
\ee
where we have used $\cT_{11} = (A+BX)^{-2}$. 
Then, using the previous relations, we obtain the expressions of the other coupling function in term of $\tilde{f}$. They read
\begin{align}
\tilde{\alpha}_2 & = \frac{B \left( A + B X\right)}{A} \tilde{f}\,, \\
\label{al3}
\cT_{33}\tilde{\alpha}_3 & = 2 \lambda_3 \tilde{X}_X \tilde{f}_{\tilde{X}}  - \gamma_3  \tilde{f}  - 2 \cJ^{-1}_g h_X   + \left(  \cT_{13} - \cT_{23} \right) A^{-1}B \left( A + B X\right) \tilde{f}\,,  \\
\cT_{44}\tilde{\alpha}_4 & = 2 \tilde{X}_X \tilde{f}_{\tilde{X}} \lambda_4 - \tilde{f}\gamma_4  +  \cT_{14} A^{-1}B \left( A + B X\right) \tilde{f} +  \cJ_g^{-1}\left( \frac{6 \tilde{f}^2_{\tilde{X}}A}{\tilde{f}\cJ_g} + 2 h_X \right)\,,
\end{align}
where we have used $\cT_{22} = (A+BX)^{-2}$, 
while the last coupling function reads
\begin{align}
\cT_{55}\tilde{\alpha}_5 & =  2\tilde{X}_{X} \tilde{f}_X \left\{ \lambda_5 - \frac{\cT_{35}}{\cT_{33}} \lambda_3 - \frac{\cT_{45}}{\cT_{44}} \lambda_4 \right\}\nonumber \\
& - \tilde{f} \left\{  \gamma_5 - \frac{\cT_{35}}{\cT_{33}} \gamma_3 - \frac{\cT_{45}}{\cT_{44}} \gamma_4 + \frac{B \left( A + B X\right)}{A} \left[ -\cT_{15}  + \cT_{25} + \frac{\cT_{35}}{\cT_{33}} \left( \cT_{13} - \cT_{23}\right) + \cT_{14} \frac{\cT_{45}}{\cT_{44}}\right] \right\} \nonumber \\
\label{condd2}
& - 2 \cJ^{-1}_g \left\{  h_{X} \left( \frac{\cT_{45}}{\cT_{44}} - \frac{\cT_{35}}{\cT_{33}} \right) + \frac{3 \tilde{f}^2_{\tilde{X}}A}{\tilde{f}\cJ_g} \frac{\cT_{45}}{\cT_{44}}\right\}\,.
\end{align}
These constraints show that, for a given choice of the potentials $A$ and $B$, the four coupling functions of the initial DHOST theory, $(\tilde{\alpha}_1, \tilde{\alpha}_2, \tilde{\alpha}_3, \tilde{\alpha}_4, \tilde{\alpha}_5 )$, can be expressed in term of the free function $\tilde{f}$. This sub-space of theories will turn out to be crucial to generate new exact solutions of DHOST theories which satisfy the observational constraints. We shall come back to this point later on.

\section{Disformal transformation of Einstein-Scalar seed solution}

\label{sec3}

In this section, we discuss the construction of new exact solutions of DHOST theories starting from a naked singularity seed solution of the massless Einstein-Scalar system. As mentioned earlier, provided the self-interacting potential is semi-positive, the no hair theorem prevents the canonical minimally coupled scalar field in GR from having asymptotically flat black hole solution with a scalar hair \cite{Agnese:1985xj, Bronnikov:2001ah, Bronnikov:2016wqp}. Preserving the minimal coupling, black holes can be obtained by relaxing asymptotic flatness \cite{Yu:2020bxd}, working with non semi-positive self-interacting potential \cite{Cadoni:2015qxa, Cadoni:2015gfa}, as well as working with a phantom field \cite{Bronnikov:2006fa, Bronnikov:1998hm, Bronnikov:1998gf}. Such phantom configuration is not an interesting starting point, since the ghost would persist in the DHOST frame. Therefore, we focus on the canonical scalar field case. The most studied example of a naked singularity solution is known as the Fischer-Janis-Newman-Winicour solution \cite{Janis:1968zz, Janis:1970kn, Virbhadra:1997ie}. Several other useful solutions describing naked singularities configurations are known for this system \cite{Reiris:2015zaa, Banijamali:2019gry, Faraoni:2018xwo, Husain:1994uj, Turimov:2018guy, Kovacs:2018yrq}, but we restrict here to the simplest spherically symmetric and static one. It is worth pointing out that even if black holes are so far the main focus in the literature, naked singularity solutions provide nevertheless interesting configurations with several useful applications. In particular, if one considers such a solution as an effective description of a compact object valid above the length scale $r_s$ at which the singularity develops, it may serve as a powerful accelerator of particles. As such, naked singularities have been used to model the highly energetic emission from the center of our galaxy such as gamma ray burst \cite{Chakrabarti:1994ig, Singh:1998vw, Patil:2011ya, Patil:2011aw, Patil:2011aa}. It is therefore an interesting geometry on its own. See \cite{Joshi:2012mk} for a review.

Below, we present this exact solution of the massless Einstein-Scalar system. Then, we generate a new family of exact solutions of DHOST obtained by the disformal transformation of this seed solution and discuss the challenge of dressing the naked singularity with a black hole horizon.

\subsection{The Janis-Newman-Winicour naked singularity}

The globally naked singularity solution of Janis-Newman-Wninicour \cite{Janis:1968zz} is described by the metric 
\begin{equation}
\label{JNW}
 d\tilde{s}^2 = \tilde{g}_{\alpha\beta} dx^{\alpha} dx^{\beta}
  =  - F^{\gamma} dt^2 + F^{-\gamma} dr^2 + r^2 F^{1-\gamma} d\Omega^2\,,
\end{equation}
where 
\be
F(r) = 1- \frac{r_s}{r}\,.
\ee
The associated scalar profile reads
\be
\label{scalarprofile}
\phi(r) = \frac{q}{r_s\sqrt{4\pi}} \log\left( 1-\frac{r_s}{r}\right)\,,
\ee
such that its kinetic energy reads
\be
\label{X}
\tilde{X}(r) = \tilde{g}^{\mu\nu} \phi_{\mu} \phi_{\nu} = \frac{q^2}{4\pi} \frac{1}{r^4} \left( 1-\frac{r_s}{r}\right)^{\gamma-2}\,.
\ee
The parameters of the solution are related to the mass of the geometry and the scalar charge, which read
\be
\gamma = \frac{2M}{r_s} \,, \quad r_s = 2 \sqrt{M^2 + q^2}\,,
\ee
such that $0 \leqslant \gamma \leqslant 1$.
Notice that when the scalar charge vanishes, i.e $q=0$, $r_s= 2M$ and $\gamma=1$, then one recovers the standard Schwarzschild vacuum geometry. Let us now briefly discuss how the JNW geometry differs from its Schwarzschild limit.
The curvature invariants are given by
\begin{align}
  \tilde{R} (r) & =  \frac{1-\gamma^2}{2} \frac{r_s^2 F(r)^{\gamma-2}}{r^4}\,,  \\
  \tilde{R}_{\mu\nu\rho\sigma} \tilde{R}^{\mu\nu\rho\sigma} (r)& = \frac{1}{12} \frac{{r_s}^2 F(r)^{2 \gamma -4 } \Delta_1(r)}{ r^8}\,, \\
  \tilde{C}_{\mu\nu\rho\sigma} \tilde{C}^{\mu\nu\rho\sigma} (r)& = \frac{1}{3} \frac{r_s^2 F(r)^{2\gamma-4}\Delta_2 (r)}{ r^8}\,,
  \end{align}
where the functions $\Delta_{1,2}(r)$ are given by
\begin{align}
\Delta_1(r) & = 4 \left[(\gamma +1) (2 \gamma +1) r_s-6 \gamma  r \right]^2 + 5 r_s^2 (1-\gamma^2)^2\,,\\
\Delta_2(r) & = \left[(1+\gamma)(1+2\gamma)r_s - 6\gamma r\right]^2\,.
\end{align}
From the power of $F(r)$ in each of the three curvature invariants, it is clear that the geometry has a curvature singularity at $r= r_s$ where each invariant blows up unless $\gamma = 1$.
The line element describes therefore the geometry for $r\in ]r_s, + \infty[$. Notice that when $\gamma=1$, the above expressions reduce as expected to the standard Schwarzschild case, the three curvature invariants are finite at $r=r_s$ and the singularity is located only at $r=0$. See \cite{Virbhadra:1995iy, Virbhadra:1998dy, Virbhadra:1998kd, Harada:2001nj} for details. 

Let us now show the absence of horizon for this geometry. As it is well known, a powerful tool to investigate the causal structure of spherically symmetric spacetimes is given by the Kodama vector. See \cite{Kodama:1979vn, Abreu:2010ru} for details on the Kodama vector in spherically symmetric spacetime. Computing the norm of the Kodama vector, we find that
\be
\tilde{\cK}_{\alpha} \tilde{\cK}^{\alpha} =  - \frac{1}{r^2 F}  \left[ r- \left( 1+ \gamma\right) \frac{r_s}{2}\right]^2\,.
\ee
It is clear that for $0\leqslant\gamma < 1$, the Kodama vector remains time-like in the whole region $r \in ] r_s, + \infty [$ and thus there is no horizon. However, for $\gamma=1$, one finds an horizon at $r=r_s$, which corresponds to the standard Schwarzschild event horizon as expected. Let us point out additionally that the scalar field satisfies the weak energy condition. Using the time-like vector $u^{\alpha} \partial_{\alpha} = \sqrt{1/ F^{\gamma}} \partial_t$ satisfying $\tilde{g}_{\alpha\beta} u^{\alpha} u^{\beta} =-1$, one has
\be
\tilde{T}_{\mu\nu} u^{\mu} u^{\nu} = \frac{1}{2} \tilde{X}(r) \geqslant 0 \qquad \forall r \geqslant r_s
\ee
It is worth pointing out that while this solution provides a simple example of a naked singularity in GR, it is nevertheless not a counterexample to the cosmic censorship conjecture which apply to dynamical collapse. Finally, let us mention that the properties of the JNW solution have been investigated in details, among which are its gravitational lensing, the motion of particle around it, as well as its stability under scalar perturbations \cite{Virbhadra:2002ju, Shaikh:2019hbm, Chowdhury:2011aa, Sadhu:2012ur}.

\subsection{A new family of exact solutions for DHOST}

Performing the disformal transformation on the JNW geometry, one obtains a new family of exact solutions of a subsector of DHOST theories. The metric of the geometries belonging to this new family reads
\be
\label{JNWdhost}
ds^2 = g_{\mu\nu}dx^{\mu}dx^{\nu} = \frac{1}{A(r)} \left\{ - F^{\gamma} dt^2 + G F^{-\gamma} dr^2 +  r^2 F^{1 - \gamma}d\Omega^2\right\}\,,
\ee
where the functions $F(r)$ and $G(r)$ are given by
\begin{align}
F(r) & = 1-\frac{r_s}{r}\,,\\
\label{g}
G(r) & = 1 - \frac{q^2}{4\pi} \frac{F^{\gamma -2}}{r^4} B\left(\phi(r), X(r) \right)\,.
\end{align}
The disformal transformation leaving the scalar field unaffected, the scalar profile is given again by its GR expression (\ref{scalarprofile}).
Using the above metric and the scalar profile, one obtains the form of the kinetic term in the DHOST frame which reads
\be
\label{XX}
X(r) = g^{\mu\nu} \phi_{\mu} \phi_{\nu} = \frac{q^2}{4\pi} \frac{A(r)}{G(r)} \frac{1}{r^4} F^{\gamma-2}\,.
\ee
For the JNW initial solution, one can then check the consistency relation (\ref{Xtilde}) between (\ref{X}) and (\ref{XX}), which is trivially satisfied. This concludes the disformal transformation of the JNW solution of the Einstein-Scalar system. Before discussing the possibility to have black hole solutions, let us point that considering perturbations on top of these new solutions might lead to pathological behavior, as some subsectors of the perturbations are known to be invariant under a general disformal mapping \cite{Motohashi:2015pra, Minamitsuji:2014waa, Tsujikawa:2015upa, Papadopoulos:2017xxx}. Hence, the instability of the seed solution due to the presence of a naked singularity would persist in the DHOST frame. 

\subsection{No-go for changing the naked singularity to a black hole horizon}

Now, we would like to investigate whether the DHOST higher order terms allow, through the free potentials $A(r)$ and $G(r)$, to modify the causal structure of the naked singularity at $r=r_s$. Under a set of assumptions that simplifies the analysis, we shall investigate if it is possible to change the asymptotically flat naked singularity in the Einstein frame to a black hole horizon in the DHOST frame.

In order to track the existence of a black hole horizon at $r=r_s$ in the DHOST frame, we need to require that
\begin{itemize}
% \item  (a) $r=r_s$ is a null hypersurface,
\item (a) the Kodama vector is timelike, null and spacelike for $r>r_s$, $r=r_s$ and $r<r_s$, respectively,
\item (b) the curvature invariants remain regular at $r=r_s$,
\item (c) the metric is Lorentzian for both $r>r_s$ and $r<r_s$ and its determinant is finite at $r=r_s$,
%\item (e) $X$ is analytic at $r=r_s$, %and it changes sign at the horizon.
\end{itemize}
and for simplicity we assume that $A$ and $G$ behave as $A \sim c_1 F^{\alpha}$ and $G \sim c_2 F^{-\beta}$ near $r=r_s$, where $c_1$, $c_2$, $\alpha$ and $\beta$ are constants\footnote{ Notice that condition (a-b) imply that $r=r_s$ is a null hypersurface since codama vector is the normal vector to surface $const = R[r] := \sqrt{g_{\theta \theta}}$.}. Rescaling the coordinates $t$ and $r$ to absorb $c_1$ and $c_2$ and neglecting subleading contributions, we have 
\begin{equation}\label{agforms}
 A = F^{\alpha}\,, \quad G = F^{-\beta}\,. 
\end{equation}

Let us begin the proof of the no-go result. 
%We start by stating that 
%the condition (a) is
%\be
%\lim_{r\to r_s} \frac{F^{\gamma}}{A} = 0\,.
%\ee
%Next, 
In order to impose the condition (a), we compute the norm of the Kodama vector as
\be
\label{kodd}
\cK_{\alpha} \cK^{\alpha}  = - r^2 \frac{F}{G} \left[ \frac{1}{r} + \frac{1-\gamma}{2} \frac{F'}{F} - \frac{A'}{2A}\right]^2\,.
\ee
We then compute the determinant of the metric (\ref{JNWdhost}) which will be useful to demand the condition (c). It reads
\be
\label{deeet}
- \text{det}(g) = \frac{G F^{2(1-\gamma)}}{A^4} r^4 \sin^2{\theta}\,.
\ee 
Finally, we can compute the three curvature invariants in order to impose the condition (b). 

In the rest of this section we assume { $A$ and $G$ behave as (\ref{agforms})}. 
%The condition (a) is then restated as
%\begin{equation}
% \gamma > \alpha\,. \label{eqn:condition(a)}
%\end{equation}
We consider the following three cases separately: 
\begin{itemize}
\item (i) $\beta \leq -1$; 
\item (ii) $-1 < \beta \leq 1$; 
\item (iii) $\beta > 1$.
\end{itemize}

For (i), it is obvious that $\cK_{\alpha} \cK^{\alpha}$ does not vanish at $r_s$ and thus the condition (a) cannot be fulfilled. In the following we thus consider the cases (ii) and (iii) only.

For (ii), the condition (a) requires that 
\be
\alpha = 1 - \gamma\;, \qquad \text{and} \qquad \beta = 0\,.
\ee 
In this case, computing the curvature invariants, one finds that the condition (b) demands that $\gamma = 1/2$ or $\gamma = 1$. The latter ($\alpha = 0$, $\beta = 0$, $\gamma = 1$) corresponds to the Schwarzschild geometry without a scalar hair and is not of our interest. On the other hand, the former ($\alpha = 1/2$, $\beta = 0$, $\gamma = 1/2$) does not satisfy the condition (c) 
%{\color{red} restated as (\ref{eqn:condition(a)}). The condition (c)} is not satisfied either
since $- \text{det}(g) \propto F^{-1}$ is negative in the region $r<r_s$. Therefore, the case (ii) does not work.

We now consider the case (iii). The condition (a) gives $\cK_{\alpha} \cK^{\alpha} \sim -(\alpha + \gamma -1)^2F^{\beta-1}/4$ near $r=r_s$ for $\alpha + \gamma \ne 1$ and $\cK_{\alpha} \cK^{\alpha} \sim -F^{\beta+1}$ near $r=r_s$ for $\alpha + \gamma = 1$. The condition (b) is equivalent to $\alpha + \gamma \geq 1$. Since the norm of the Kodama vector has to change sign at the horizon, it implies that $\beta = 2n$ where $n \in \mathbb{N}$. The condition (c) is then restated as $2\alpha + \gamma = m$ ($m=0,\pm 1, \pm 2, \cdots$) and $1-m-n \geqslant0$ so that $- \text{det}(g) \propto F^{2(1-m-n)}$ is positive for both $r>r_s$ and $r<r_s$ and is finite at $r=r_s$. Also, excluding the trivial Schwarzschild case ($\gamma=1$), it follows that $0 \leq \gamma < 1$. In summary the conditions (a)-(c) are satisfied if and only if
\begin{equation}
 \beta = 2n\,, \quad (n=1,2,\cdots\,) \,, 
\end{equation}
and 
\begin{equation}
 2\alpha + \gamma = m\,, \quad 1- m -n \geq 0\,,  \quad \alpha + \gamma \geq 1\,, \quad 0 \leq \gamma < 1\,,
%  \quad (m=0,\pm 1, \pm 2, \cdots )\,. \label{eqn:alphagamma-case(iii)}
\end{equation}
Expressing $\gamma$ in term of $\alpha$ and $m$, the above conditions can be recast as
%The conditions (\ref{eqn:alphagamma-case(iii)}) for $\alpha$ and $\gamma$ are simultaneously satisfied only with $m=2$ and thus is reduced to 
\begin{eqnarray}
\alpha \geqslant 0\,, \quad m- \alpha \geqslant 1\;, \quad \frac{m-1}{2} <\alpha \leq \frac{m}{2}
% \frac{1}{2} < \alpha - \frac{m}{2} < \frac{2}{3}\,, \quad \gamma = 2 - 2\alpha\,. 
\end{eqnarray}
for which there is no admissible solutions satisfying also $\beta =2 n >1$.

%This implies that $\alpha + \gamma > 1$ and thus $\cK_{\alpha} \cK^{\alpha} \sim -(\alpha + \gamma -1)^2F^{2n-1}/4$ near $r=r_s$. Hence, in the DHOST frame, $r=r_s$ represents a non-degenerate horizon for $n=1$ and a degenerate horizon for $n=2,3,\cdots$. However, we have
%\begin{equation}
% X = \frac{q^2}{4\pi}\frac{1}{r^4}F^{2 n + m - \alpha - 2}\,,
%\end{equation}
%and $2n-\alpha = -\alpha < 0$. Therefore, the condition (d) is not satisfied. This conclude the proof that the hypersurface located at $r_s$ can not be turn into a black hole horizon within the DHOST frame.

Let us now show that this conclusion extends to any horizon located $r_{\ast} \neq r_s$. Considering the norm of the Kodama vector (\ref{kodd}), it is easy to see that the spacetime admits an horizon at $r_{\ast} \neq r_s$ provided the function $G$ changes sign, since inside the trapped region, the Kodama vector has to be spacelike. However, if $G$ changes sign at $r_{\ast}$, then the determinant (\ref{deeet}) will change sign too, preventing the spacetime from being Lorentzian both outside and inside such horizon. In the end, this prevents one to dress the naked singularity with a well defined black hole horizon at some radius $r_{\ast} > r_s$, completing therefore the no go result.

To conclude, we have shown that starting from the Janis-Newman-Winicour naked singularity as a seed solution, one can indeed obtain a new family of exact solutions of the quadratic DHOST theories. However, there is no admissible choices for the free potentials $A(r)$ and $B(r)$ such that this new family of exact solutions contains asymptotically flat black hole geometries. As the tractable aymptotically flat spherically symmetric and static solution of the canonical Einstein-Scalar system represent naked singularity, it seems hopeless to use these kind of geometries as seed solution to generate new exact asymptotically flat black hole solutions for DHOST theories\footnote{Exact solutions of the Einstein-Scalar system presented in \cite{Cadoni:2015qxa, Cadoni:2015gfa} might provide another type of seed solutions, but they turn out to be quite involved. Another possibility would be to consider for example the exact solution presented recently in \cite{Yu:2020bxd}.}. Nevertheless, we emphasize that naked singularity geometries can still provide interesting geometries, in modeling high energy phenomena, such as gamma ray burst emission. See \cite{Chakrabarti:1994ig, Singh:1998vw, Patil:2011ya, Patil:2011aw, Patil:2011aa}. Let us now discuss the disformal transformation of exact black hole solution of DHOST theories.

\section{Generating black hole solutions beyond the stealth sector}

\label{sec4}

In this section, we consider the known exact black hole solutions of quadratic DHOST theories. We shall focus on two different type of solutions : stealth and non-stealth configurations.

The well known stealth Schwarzschild solution associated to a constant kinetic term was considered in several works \cite{Babichev:2012re, Anabalon:2013oea, Babichev:2013cya, Charmousis:2014zaa, Kobayashi:2014eva, Babichev:2016kdt, Motohashi:2018wdq, Minamitsuji:2018vuw} and provides the simplest solution for such higher order theories. Conditions on the coupling functions of the DHOST Lagrangian for admitting such stealth black hole solutions have been derived in \cite{BenAchour:2018dap, Motohashi:2019sen}. It provides the simplest seed solution in the context of DHOST theories.  As for non-stealth black hole solution, a useful example was found in \cite{Babichev:2017guv}. It is associated to a non-constant kinetic term and therefore provides a complementary seed solution to explore the solution generating method.

In the following, we shall review these two solutions and the subset of theories which admit them as solutions. Then, we will present the general construction of new solutions starting from them. This will allow us to present the advantages and limitations of each seed solutions to generate new ones and give a complete picture of the potential application of our solution generating method.

\subsection{Seed black hole solutions in DHOST theories}

\subsubsection{Schwarzschild stealth solution}

\label{sec2.1}
First, let us recall the general construction of stealth solutions in DHOST theories. 
A stealth black hole solution corresponds to a metric which is a vacuum solution of General Relativity, and which admits a non-trivial scalar filed with a vanishing energy-momentum tensor. In general, such stealth solutions are known with a constant kinetic term, although example with a time-dependent kinetic term exist \cite{Minamitsuji:2018vuw}. 

It was recently pointed out in \cite{deRham:2019gha} that a large class of stealth solutions in DHOST theories have a singular effective metric for linear perturbations and thus suffer from strong coupling. On the other hand, it has been known in the context of ghost condensation~\cite{ArkaniHamed:2003uy,ArkaniHamed:2003uz} that an approximately stealth black hole solution exists and that linear perturbations around it are under a good theoretical control~\cite{Mukohyama:2005rw,Mukohyama:2009rk,Mukohyama:2009um}. The approximately stealth solution is based on the expansion with respect to coefficients of higher dimensional operators and it reduces to the exact stealth solution at the zero-th order in the expansion. Those higher dimensional operators do not significantly change the background geometry from the GR solution but make the dispersion relation of linear perturbations non-singular. This approximately stealth solution can be extended to more general setups and thus provides a resolution of the strong coupling problem of the stealth solution in DHOST theories, provided that the derivative of the scalar field is timelike~\cite{Motohashi:2019ymr}. In the present paper, whenever we study stealth solutions in DHOST theories and their disformal transformations, we have in mind this resolution of the strong coupling issue. This in particular means that the derivative of the scalar field should be timelike.

As an example of such stealth black hole solutions, consider the simple Schwarzschild vacuum geometry
\be
ds^2 = - F(r) dt^2 + F(r)^{-1} dt^2 + r^2 d\Omega^2\,, \quad F(r) = \left( 1 - \frac{r_s}{r} \right)\,.
\ee
Under suitable conditions derived in \cite{BenAchour:2018dap, Motohashi:2019sen}, quadratic DHOST theories admit such geometry as an exact solution associated to the scalar profile
\be
\label{tdscal}
\phi(t,r) = q t + \psi(r)
\ee
while its kinetic term remains constant, such that
\be
\tilde{X}_{\circ} = \tilde{g}^{\mu\nu} \phi_{\mu} \phi_{\nu}  = - \frac{q^2}{F} + F \left( \psi'\right)^2\,.
\ee
As mentioned above, the resolution of the strong coupling issue requires that $\partial_{\mu}\phi$ be timelike. We thus suppose that
\begin{equation}
  -q^2 \leq \tilde{X}_{\circ} < 0\,,
\end{equation}
where the first inequality follows from the reality of $\psi'$. Let us introduce for simplicity the notation
\be
\label{def}
Z = \tilde{X}_{\circ} F+ q^2 \;, \qquad Z_{\circ} = \tilde{X}_{\circ}+q^2\,.
\ee
This allows us to fix the form of the spatial gradient of the scalar field to 
\be
\psi'(r) = \pm \sqrt{\frac{Z}{F^2}}\,. 
\ee
%Although the gradient blows up at $r=r_s$, one can re-absorb this coordinate-dependent behavior by a suitable change of coordinate when working in the suitable branch. 
When $r\rightarrow +\infty$, it asymptotes to $\psi' \sim \pm \sqrt{Z_{\circ}}$.
Integrating this equation, one obtains the expression for the function $\psi(r)$ which reads
\begin{align}
 \psi(r) & = \psi_{\circ} \pm \left\{ r\sqrt{Z}   + \frac{\left( 2q^2 + \tilde{X}_{\circ}\right)r_s}{\sqrt{Z_{\circ}}} \log{\left[ \sqrt{r Z_{\circ}} \left( \sqrt{Z_{\circ}} + \sqrt{Z} \right)\right]}   + q r_s \ln \left|\frac{q-\sqrt{Z}}{q+\sqrt{Z}}\right| \right\}\,.
\end{align}
This profile blows up going from $\phi \rightarrow + \infty$ for $r\rightarrow + \infty$ to $\phi \rightarrow - \infty$ at $r=r_s$. The assumption of the constant kinetic term fully fixes the form of the linear time-dependent scalar profile, up to the choice of the sign of $\psi'$ and the three integration constants $\tilde{X}_{\circ}$, $q$ and $\psi_{\circ}$.

This stealth Schwarzschild geometry is a solution of shift symmetric quadratic DHOST theories provided that the following constraint holds
\begin{align}
\label{cond0}
\tilde{P}= \tilde{P}_{\tilde{X}} & = \tilde{Q}_{\tilde{X}} =0\\
\label{cond1}
 \tilde{\alpha}_{1} = \tilde{\alpha}_2 & = \tilde{\alpha}_{1\tilde{X}} + \tilde{\alpha}_{2\tilde{X}} \\
\label{cond2}
&  = Z_{\circ}\left( 2\tilde{\alpha}_{1\tilde{X}} + \tilde{\alpha}_3\right) = 0 
\end{align}
when evaluated at $\tilde{X}_{\circ}$. See \cite{BenAchour:2018dap, Motohashi:2019sen} for details. Let us now check whether these conditions can be satisfied by a DHOST theory belonging to the sub-sector we have identified in Section-\ref{sec1.3}. First, the condition $\tilde{P}= \tilde{P}_{\tilde{X}}=\tilde{Q}_{\tilde{X}} =0$ at $\tilde{X}_{\circ}$ does not have any impact on the analysis in Section-\ref{sec1.3} and can be imposed independently. Moreover, we observe that the conditions (\ref{cond1}) at $\tilde{X}_{\circ}$ imply that $\tilde{f} = \tilde{f}_{\tilde{X}} =0$ at $\tilde{X}_{\circ}$, which makes the rest of the conditions consistent with the relations (\ref{condd1}) to (\ref{condd2}). 

Noticed that the full set of conditions for non-shift symmetric quadratic DHOST theories for admitting stealth Schwarzschild black hole solution is known only for the sub-case $\alpha_1 =\alpha_2=0$. From the conditions found in \cite{BenAchour:2018dap}, it appears that even in this sub-case, conditions between the coupling functions will be needed beyond shift symmetry, restricting the allowed form of the disformal potentials $A$ and $B$ in general.

As a consequence, we known that i) the sub-sector identified in Section-\ref{sec1.3} is mapped back, by construction, to the sub-sector consistent with the observational constraints and ii) that it admits stealth Schwarzschild solution. Starting from this sub-sector of shift symmetric theories ensures us to generate new exact solutions of the viable sub-sector of DHOST theories, i.e those which satisfy $c_g=c$ and which avoid the decay of gravitons to scalar perturbations. Let us now describe the second type of seed solution which corresponds to a non-stealth black hole.

\subsubsection{Reissner-Nordstrom non-stealth solution}
\label{subsubsec:RNnonstealth}

Another useful black hole solution of shift symmetric quadratic Horndeski theory was obtained in \cite{Babichev:2017guv}. The Lagrangian of this theory reads
\be
\label{lagcris}
\tilde{\cL} = \tilde{P}(\tilde{X}) + \tilde{f}(\tilde{X}) \tilde{\cR} -2 \tilde{f}_{\tilde{X}} \left[ \left( \tilde{\Box} \phi\right)^2 - \phi_{\mu\nu} \phi^{\mu\nu}\right]\,,
\ee
where the coupling functions are given by
\be
\label{coupcris}
\tilde{P}(\tilde{X}) = -\frac{1}{2} \eta \tilde{X} \,, \quad \tilde{f}(\tilde{X}) = \zeta + \beta \sqrt{ \frac{1}{2}\tilde{X}}\,,
\ee
where $\zeta >0$~\footnote{In \cite{Babichev:2017guv}, $\tilde{X}$ is denoted as $-2X$}. Therefore, the standard canonical action is recovered for $ \eta > 0$. We shall focus only on this range of parameters. 
This theory admits an exact black hole solution of the form
\begin{align}
ds^2 = - F(r) dr^2 + \frac{dr^2}{F(r)} + r^2 d\Omega^2\,,
\end{align}
with
\be
F(r) = 1 - \frac{\mu}{r} + \frac{\eta}{4\zeta r} \int dr r^2 \tilde{X}\,.
\ee
The scalar profile is time-dependent and has the same form as in (\ref{tdscal}) although with the different radial contribution $\psi(r)$. The associated kinetic term is given by the equation
\be
\label{fonc44}
\left( \sqrt{\frac{1}{2}\tilde{X}}\right)^2 \left( 1- \frac{\eta}{\beta} r^2 \sqrt{\frac{1}{2}\tilde{X}}\right) = - \frac{q^2}{2}\,,
\ee
The discriminant being negative\footnote{ The discriminant reads
\be
\Delta = - \frac{q^2}{2} \left( 4 + \frac{27 \eta^2}{\beta^2} r^2\right) <0
\ee
which implies that $\sqrt{\tilde{X}/2}$ has two conjugated complex solutions and only one real solution.} for any $r$, there is only one single real root for $\sqrt{\tilde{X}/2}$ which is given by
\begin{align}
\label{XX}
\sqrt{\frac{1}{2}\tilde{X}} & = \frac{\beta}{3\eta r^2} \left( 1 +  K ^{1/3} + K^{-1/3}\right)\,,\\
 K & = \frac{1}{2} \left[ 2 + \frac{27 q^2 \eta^2 r^4}{2\beta^2} - \sqrt{- 4 + \left( \frac{27 q^2 \eta^2 r^4}{2\beta^2} + 2\right)^2} \right]\,.
\end{align}
The function $K(r)$ is always positive, never vanishes and remain finite everywhere\footnote{In (3.17) of \cite{Babichev:2017guv}, in order for its right hand side to be real, it should be understood that $(-1)^{1/3}=-1$ instead of $(-1)^{1/3}=(1+i\sqrt{3})/2$. Also, $K$ is denoted as $-A$ in \cite{Babichev:2017guv}.} and the kinetic term $\tilde{X} = \tilde{g}^{\mu\nu} \partial_{\mu} \phi \partial_{\nu} \phi>0$ for all $r$ for such configuration. This shall have important consequence in the following. Using this solution, one can show that the geometry admits a single horizon located at a radius that we shall denote $r_{\circ}$ in what follows. Moreover, its asymptotic behavior is given by
\be
F(r) \sim \frac{3\eta}{10\zeta} \left( \frac{q^2 \beta}{2\eta}\right)^{2/3}  r^{2/3}\;, \qquad \text{when} \qquad r \rightarrow + \infty\,.
\ee
The radial contribution to the scalar profile can be obtained by integrating 
\be
F^2 \left(\phi'\right)^2 = q^2 + \frac{2\beta^2}{9\eta^2 r^4} F \left( 1+ K^{1/3} + K^{-1/3}\right)\,.
\ee
The crucial point is that this exact black hole solution, although quite complicated, provides one of the few known examples which describe a non-stealth black hole geometry, with a non constant kinetic term and a time-dependent scalar profile. From the point of view of the three properties stated earlier in introduction, this solution satisfies all the three. As we shall see in the next section, it is therefore not a surprise that it serves as a seed solution allowing for the most general modifications when constructing new black hole solutions using a disformal mapping.  Before presenting this construction, let us present its static limit. 

In the static case, where we have $q=0$, the solution becomes
\be
\label{fonc1}
F(r) = 1 - \frac{\mu}{r} - \frac{\beta^2}{2\zeta \eta r^2}\,.
\ee
Notice that because $\zeta >0$ and $\eta >0$, the metric has the same form as a Reissner-Nordstrom metric with an imaginary electric-like charge. 
The associated scalar profile is given by
\begin{align}
\label{fonc2}
\phi(r) & = \pm 2 \sqrt{\frac{\zeta}{\eta}} \left\{ \text{Arctan} \left[ \frac{\beta^2 + \zeta \eta \mu r}{\beta \sqrt{2\zeta \eta r \left( r-\mu\right) -\beta^2}}\right]  - \text{Arctan} \left( \frac{\mu}{\beta} \sqrt{\frac{\zeta \eta}{2}}\right) \right\}
\end{align}
when $ \beta >0$ and $\eta > 0$, while it is given by 
\begin{align}
\label{fonc3}
\phi(r) & = \pm 2 \sqrt{\frac{\zeta}{\eta}} \left\{ \text{Arcth} \left[ \frac{\beta^2 + \zeta \eta \mu r}{\beta \sqrt{2\zeta \eta r \left( r-\mu\right) -\beta^2}}\right]  - \text{Arcth} \left( \frac{\mu}{\beta} \sqrt{\frac{\zeta \eta}{2}}\right) \right\}
\end{align}
when $ \beta <0$ and $\eta < 0$. 
Finally, the kinetic term of the scalar field is not constant and decays as $r^{-4}$, such that
\be
\label{fonc4}
\tilde{X} = \frac{2\beta^2}{\eta^2 r^4}\,.
\ee
This exact solution provides therefore an interesting example of a non-stealth configuration with a rather simple profile for the kinetic term. Indeed, when $q\neq 0$, one can easily write the radius $r$ in term of $\tilde{X}$ using (\ref{fonc44}), while when $q=0$, the relation (\ref{fonc4}) is straightforward to invert. This ensures that one can realize any $r$-dependence of the potentials $A$ and $B$ through their dependence on $\tilde{X}$. Finally, when working with this seed solution, we shall introduce the generation of (\ref{def}) which reads
\be
Z = \tilde{X} F + q^2\,.
\ee

Before presenting the construction of new black hole solutions, we mention here that we shall not impose the condition for the theory (\ref{lagcris})-(\ref{coupcris}) to be mapped back to the sub-sector of theories summarized in section~\ref{subsec:consraints}. Nonetheless, a black hole solution with $q \ne 0$ may still be observationally viable if $\phi \to const$ at cosmological distances. Depending on the black hole formation scenario, this may be the case. For example, starting with a cosmological solution with $\phi = const$, a gravitational collapse of a compact object may induce a non-trivial evolution of the scalar field and consequently a scalar hair may develop while a black hole forms. The scalar field in the vicinity of the black hole may settle to a stationary configuration within a finite relaxation time but it should take a long time for the scalar hair to propagate to the cosmological distances. If this is the case then the black hole with a $q\ne 0$ scalar hair may be consistently embedded in a cosmological background with $\phi = const$. In this case the phenomenological constraints summarized in section~\ref{subsec:consraints} does not need to be imposed. It is certainly worthwhile investigating the formation of a black hole in DHOST theories to see if this may happen or not.

\subsection{Constructing new hairy black hole solutions}

\label{sec2.2}

As explained above, each of the three different subclasses of DHOST theories corresponding to a given degeneracy condition is stable under a general disformal transformation.  Therefore, one natural way to generate new solutions for each class is to perform a disformal mapping on the known exact solution for a given sub-sector of DHOST theories. The goal of this section is to discuss the outcome of this procedure. 

As a first step, we intend to present the outcome of the disformal transformation on a general metric ansatz. Consider therefore the seed metric given by
\begin{align}
\label{seedm}
d\tilde{s}^2 & = \tilde{g}_{\mu\nu} dx^{\mu} dx^{\nu}  = - F(r) dr^2 + \frac{dr^2}{F(r)} + r^2 d\Omega^2\,,
\end{align}
which encompasses the two exact solutions discussed previously. Moreover, we work with the general scalar profile
\be
\label{seedscal}
\phi(t,r) = qt + \psi(r)\,.
\ee
Performing a disformal transformation on the general metric (\ref{seedm}) associated to the profile (\ref{seedscal}), we obtain the new exact solution
\be
ds^2 = g_{\mu\nu}dx^{\mu}dx^{\nu} =  A^{-1} \left[ - F G_1 dt^2 - 2 N_r dt dr + \frac{G_2}{F} dr^2 + r^2 d\Omega^2\right]\,,
\ee
where the new functions $G_1$, $G_2$ as well as the shift are related to the free potential $B(r)$ through
\begin{align} 
\label{eqn:G1G2Nr-B}
G_1(r) & = 1 + \frac{q^2 B(r) }{F}\,, \\
\label{eqn:G1G2Nr-BB}
 G_2 (r) & = 1 - \frac{Z B(r) }{F}\,, \\
 \label{eqn:G1G2Nr-BBB}
 N_r & = \pm  \frac{qB(r)}{F} \sqrt{Z }\,,
\end{align}
where $Z(r) = \tilde{X}(r) F(r) + q^2$ in general. The plus and minus signs for $N_r$ correspond to a scalar profile that regularly penetrates the black hole horizon and the white hole horizon, respectively. We can now apply this general construction to the two seed solutions reviewed in the previous section.

\subsubsection{Starting from a stealth solution}

\label{secstealth}

As a first example we consider the Schwarzschild seed solution with a constant kinetic term. As a first remark, notice that since we have assumed that $A:=A(X)$ and $B:= B(X)$ (otherwise they would become time-dependent for $q\neq0$), the right hand side of (\ref{Xtilde}) is a function of $X$ only. Hence, whenever (\ref{Xtilde}) is invertible with respect to $X$, $X$ can be expressed as a function of $\tilde{X}$, which is set to be the constant $\tilde{X_{\circ}}$ for the stealth seed solution considered here. Therefore, restricting to invertible disformal transformations implies that $X$, $A$ and $B$ are constant
\be
X = X_{\circ} \;, \qquad A = A_{\circ}\;, \qquad B = B_{\circ}\,.
\ee
From this, it is straightforward to see that 
\be
X_{\circ} 
 = \frac{A_{\circ}\tilde{X}_{\circ}}{1-B_{\circ} \tilde{X}_{\circ}}\,.
 \ee
 Moreover, the form of the $G_2(r)$ modification is fixed to
\be
G_2 (r)= 1 - \frac{Z(r) B_{\circ} }{F(r)}
\ee
Therefore, under the above assumptions, the metric we obtain from the stealth Schwarzschild black hole as a consequence of the disformal transformation reads
\begin{align}
\label{newwsol}
ds^2&  =  \frac{1}{A_{\circ}}\left\{ - F \left( 1 + \frac{q^2 B_{\circ} }{F}   \right) dt^2   \mp \frac{2 q  \sqrt{Z} B_{\circ} }{F} dt dr + F^{-1} \left( 1 - \frac{ZB_{\circ} }{F}\right) dr^2 + r^2 d\Omega^2 \right\}\,.
\end{align}
Notice that this geometry exhibits a deficit solid angle whenever $Z_{\circ}B_{\circ} \ne 0$. Indeed, the area radius is $r/\sqrt{A_{\circ}}$ and the $rr$-component of the the metric in the region $r\rightarrow + \infty$ becomes $g_{rr} \sim  ( 1 - Z_{\circ}B_{\circ})/A_{\circ}$. The norm of the Kodama vector is given by
\be
\cK_{\alpha} \cK^{\alpha} = - \frac{r+q^2B_{\circ}r-r_s}{r(1-\tilde{X}_{\circ}B_{\circ})}\,,
\ee
which implies that the horizon is located at
\be
r_{\ast} = \frac{r_s}{1+ q^2 B_{\circ}}\,. 
\ee
Depending on the sign of $B_{\circ}$, the new horizon radius can be either greater of smaller than $r_s$. 
Computing moreover the Misner-Sharp mass, we obtain
\be
M_{\text{MS}} = \frac{r_s - Z_{\circ}B_{\circ}r}{2\sqrt{A_{\circ}}(1-\tilde{X}_{\circ}B_{\circ})}\,.
\ee
This shows that the above solution reduces to the standard Schwarzschild only when $B_{\circ}Z_{\circ}=0$, i.e. when either $B_{\circ}=0$ or $\tilde{X}_{\circ} = - q^2$. 
Finally, computing the determinant of the metric, it reads
\be
 \text{det}(g) = -r^4 \sin ^2(\theta) \left(1-B_{\circ} \tilde{X}_{\circ} \right)A_{\circ}^{-4}\,,
\ee
which keeps the same sign in the whole spacetime. Let us summarize the outcome of the disformal transformation of the standard stealth Schwarzschild solution. Assuming that $-q^2\leq \tilde{X}_{\circ} < 0$ and $B_{\circ}\ne 0$, one can classify the solutions after the disformal transformation into the following two cases. 
\begin{itemize}
 \item Case 1: $-q^2 < \tilde{X}_{\circ} < 0$. This provides a solution which describe an asymptotically locally flat non-stealth black hole with a deficit solid angle. The deficit solid angle is traced back to the non-zero value of $\psi'$ at $r=\infty$, while the non-stealth character of the solution can be recognized from the non-trivial $r$-dependence of its Misner-Sharp energy. 
 \item Case 2: $-q^2 = \tilde{X}_{\circ} < 0$. In this case, the solution after the transformation is stealth and corresponds to a Schwarzschild black hole with a shifted mass and horizon radius. 
\end{itemize}

Finally, let us point that such disformal transformation of the Schwarzschild stealth solution provides a straightforward way to build minimal hairy deformation of vacuum GR solution which are exact solution of a DHOST theory. Up to now, such minimal deformation of the vacuum solutions of GR have been introduced for phenomenological investigation and remain ad hoc \cite{McManus:2019ulj}. Using the new exact solution (\ref{newwsol}), and performing a linear expansion in term of the parameters $A_{\circ}$ and $B_{\circ}$, our solution generating method allows instead to endow these phenomenologically interesting geometries within the effective approach of DHOST theories. Consider therefore infinitesimal version of the disformal transformation used above where
\be
A_{\circ}=1+\epsilon_1\;, \qquad B_{\circ}=\epsilon_2\;,  %\tilde{g}_{\mu\nu} = \left( 1+ \epsilon_1 \right) g_{\mu\nu} + \epsilon_2 \partial_{\mu}\phi \partial_{\nu}\phi\;,
 \qquad
 \text{with} \qquad |\epsilon_1|, \; |\epsilon_2| \ll 1\,.
\ee
Keeping only the leading terms in $\epsilon_1$ and $\epsilon_2$, one can generate using the disformal transformation a new exact solutions of DHOST theories of the form
\begin{align}
 ds^2 & = - \left[ (1-\epsilon_1)\left(1-\frac{r_s}{r}\right) + q^2\epsilon_2 + \cO (\epsilon_{1,2}^2) \right] dt^2  \mp 2 q \epsilon_2 \sqrt{\tilde{X}_{\circ} + q^2 - \frac{\tilde{X}_{\circ} r_s}{r}} \left( 1-\frac{r_s}{r}\right)^{-1} dt dr \nonumber  \\
 &  + \left( \left( 1  - \epsilon_1 - \tilde{X}_{\circ} \epsilon_2 \right) \left( 1-\frac{r_s}{r}\right)^{-1}  - q^2 \epsilon_2  \left( 1-\frac{r_s}{r}\right)^{-2}  + \cO(\epsilon_{1,2}^2)\right) dr^2  + \left( 1-\epsilon_1 + \cO(\epsilon_1^2)\right) r^2 d\Omega^2
\end{align}
We can now analyse the properties of this geometry. First of all, the horizon structure of this geometry is given by
\begin{equation}
\label{kodeps}
 \cK_{\alpha} \cK^{\alpha} (r_{\ast})  =
 \frac{(1+q^2\epsilon_2)r_{\ast}-r_s}{r(1-\tilde{X}_{\circ}\epsilon_2)}  + \cO(\epsilon^2) = 0\,,
\end{equation}
such that the position of the horizon is slightly shifted and reads
\be
r_{\ast} \simeq \left( 1- q^2 \epsilon_2 \right) r_s + \cO(\epsilon^2) \,.
\ee
From (\ref{kodeps}), it is straightforward to see that the Misner-Sharp mass,
\begin{equation}
M_{\text{MS}} = \frac{r_s-Z_{\circ}\epsilon_2r}{2\sqrt{1+\epsilon_1}(1-\tilde{X}_{\circ}\epsilon_2)} + \cO(\epsilon^2)\,,
\end{equation}
is not constant if $Z_{\circ}\epsilon_2 \ne 0$, where $Z_{\circ}$ is defined in (\ref{def}). Hence, the Misner-Sharp energy does not coincide with the horizon radius divided by 2, as it is the case for any stealth Schwarzschild solution, and therefore, the new geometry is not stealth in general.

We can now compute the curvature invariants for this new solutions. At leading order, they coincide with the Schwarzschild's ones, such that
\begin{align}
\cR & = -\frac{2 \epsilon_2 \left(q^2+\tilde{X}_0 \right)}{r^2} + \cO(\epsilon^2)\,, \\
\cR^{\mu\nu\rho\sigma} \cR_{\mu\nu\rho\sigma} = \cC^{\mu\nu\rho\sigma} \cC_{\mu\nu\rho\sigma} & = \frac{12 r_s^2}{r^6}  -\frac{8 r_s \left(r \epsilon_2 \left(q^2+\tilde{X}_0 \right)-3 r_s (\tilde{X}_0 \epsilon_2+ \epsilon_1)\right)}{r^6} + \cO(\epsilon^2)\,.
\end{align}
Finally, computing the kinetic term, one obtains that it remains constant at all order. The leading term reads
\be
X_{\circ} \simeq  \left( 1 + \epsilon_1 + \epsilon_2 \tilde{X}_{\circ} \right) \tilde{X}_{\circ} + \cO(\epsilon^2)\,.
\ee 
This concludes the construction of the minimally modified Schwarzschild black hole in the context of DHOST theories. We can now turn to the disformal transformation of non-stealth seed solution of DHOST theories.

\subsubsection{Starting from a non-stealth solution}

Now let us consider the non-stealth seed solution. %A crucial difference with the previous case is that the non-stealth solution is associated to non-constant, radially dependent kinetic term together with a time-dependent scalar profile. As we are going to see, this opens up the possibility to choose more freely the disformal potential.
%Using (\ref{eqn:G1G2Nr-B}), (\ref{eqn:G1G2Nr-B}) and (\ref{eqn:G1G2Nr-BBB}), 
Consistency imposes that the function $G_1$ and the shift $N_r$ are related to the free function $G_2$ through
\begin{align}
\label{g1}
G_1(r) & = 1 + \frac{q^2 \left( 1-G_2\right) }{Z }\,,  \\
N_r (r) & =  \pm \frac{q \left( 1-G_2\right)}{\sqrt{Z  }}\,.
\end{align}
Therefore, the freedom is encoded in the functions $A(r)$ and $G_2(r)$. Using the above relations, we obtain finally the new exact solution of DHOST theories given by
\begin{align}
\label{newsol}
ds^2&  = A^{-1} \left[  - F \left( 1 + \frac{q^2  \left( 1-G_2 \right)}{Z } \right) dt^2  \mp \frac{2 q   \left( 1-G_2\right)}{\sqrt{Z }} dt dr + \frac{G_2}{F} dr^2 + r^2 d\Omega^2 \right]\,.
\end{align}
This new class of geometries provide, by construction, a new family of exact solutions for quadratic DHOST theories. Let us now compute the profile of the kinetic term in this new geometry. It reads 
\begin{align}
X & = \frac{A \tilde{X}}{1- \tilde{X} B} = \frac{ A \tilde{X} Z }{\tilde{X} FG_2+q^2}\,.
\end{align}
This concludes the presentation of this new family of exact solutions of quadratic DHOST theories. We now can turn to the analysis of its properties. The condition for asymptotic flatness is easily obtained as
\be
\lim_{r\to\infty} A (r) = A_{\infty}\,, \quad \lim_{r\to\infty} G_2(r) = 1\,,
\ee
where $A_{\infty}$ is a positive constant. Let us now compute the norm of the Kodama vector and the determinant of the metric. They reads
  \begin{align}
\label{kod4}
\cK^{\alpha} \cK_{\alpha} & = - \frac{F \left[r  A' +2 A \right]^2 \left(q^2 (1-G_2) + Z\right)}{4 \left(\tilde{X} F G_2+q^2\right) A^2}\,,\\
\label{det4}
 \det( g) & = -\frac{r^4 \sin ^2(\theta) \left(\tilde{X} F G_2+q^2\right)}{A^4 Z}  = - \frac{r^4 \sin^2{\theta}}{A^3} \frac{\tilde{X}}{X}\,.
  \end{align}
Using (\ref{kod4}) and (\ref{det4}), the necessary conditions for having a well defined modified geometry, with possibly new horizons, while preserving at the same time the sign of the determinant, are given by
\be
\label{choice0}
0<|A(r)| < \infty\,, \qquad \text{and} \qquad \forall r \geq r_s,\; |G_2(r)| < \infty\,, 
\ee
and $q^2 >0$. 
As we are going to see now, preserving the determinant does not prevent from adding additional horizons to the seed geometry when $q\neq 0$. However, this is no longer true in the static case $q=0$.

Moreover, in order to have a well defined hairy black hole solution, it is required that the kinetic term of the scalar field $X$ remains regular on any horizons. It is straightforward to see that the behavior of $X$ is related to the behavior of the determinant $ \det( g) $ through (\ref{det4}). Since $\tilde{X}$ is always positive, then provided $ \det( g)$ does not vanish on any horizons, the kinetic term $X$ remains also regular on these horizons.

Consider therefore the case with $q\neq0$, and let us introduce a choice of the potentials $A(r)$ and $G(r)$ which satisfies the above conditions and generate a new black hole horizon. One interesting example is given by the following potentials
\be
\label{choice}
A(r)=1\,, \qquad \text{and} \qquad G_2(r) = \left(1 - \frac{r_{\ast}}{r}\right)^{2}\,,
\ee
which introduce a new arbitrary scale $r_{\ast}$. 
Depending on the value of $r_{\ast}$, one can introduce a new horizon on top on the one pre-existing at $r_{\circ}$ in the seed solution. This is made possible because the norm of the Kodama vector $\cK^{\alpha} \cK_{\alpha} =0$ can now have several roots. However, this modification introduces new singularities, which nevertheless remain hidden inside the inner horizon. The metric being now quite involved, we investigate the new structure of the solution only numerically. To this end, we shall plot the norm of the Kodama vector, the determinant and the $00$-component of the new metric as well as its Ricci scalar for the new solution for two set of the parameters $\left( \eta, \zeta, \beta, \mu, r_{\ast} \right)$. In what follows, we work with $\eta / (2\zeta) = 10^{-4}$ and $\mu = 1$ such that $r_{\circ}\sim 1$. Moreover, we set  $r_{\ast}/\mu = 4$. The properties of the seed solution are depicted in yellow, while the properties of the modified solution are depicted in blue.

\begin{figure}[h!]
  \centering
  \includegraphics[scale=0.5]{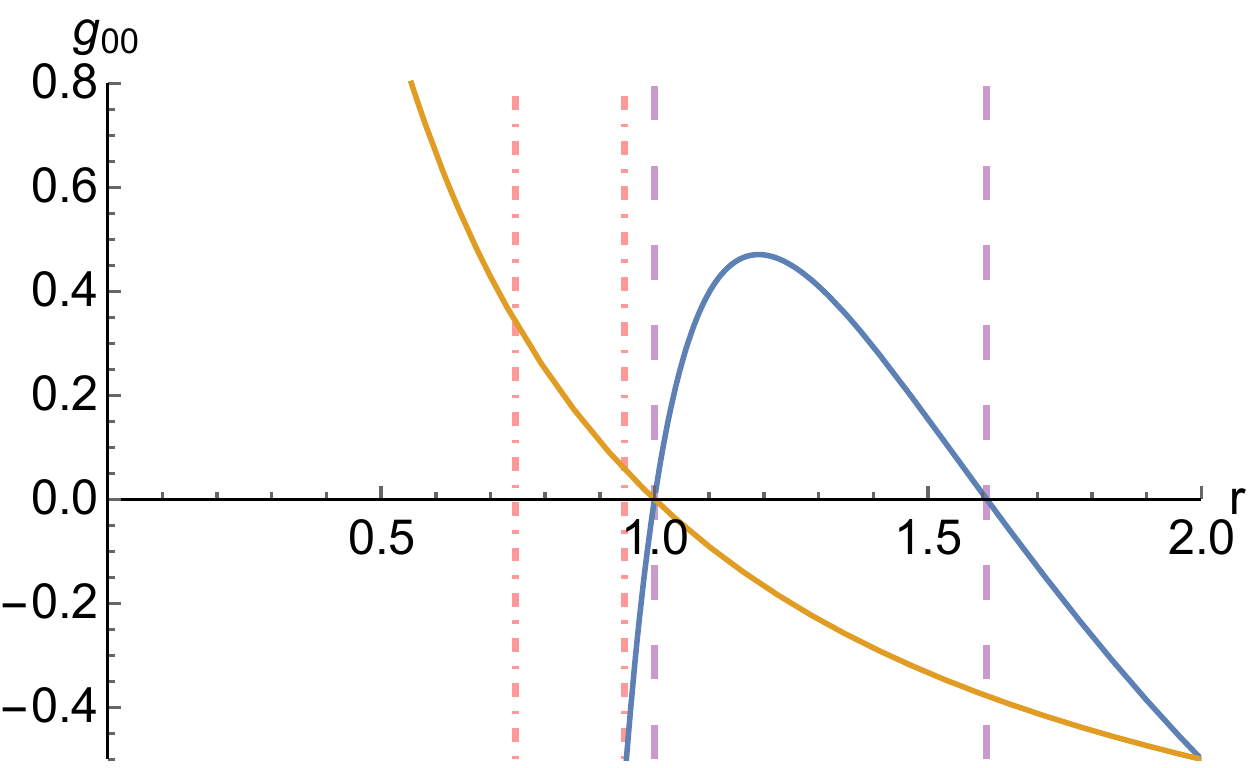}
 \; \includegraphics[scale=0.5]{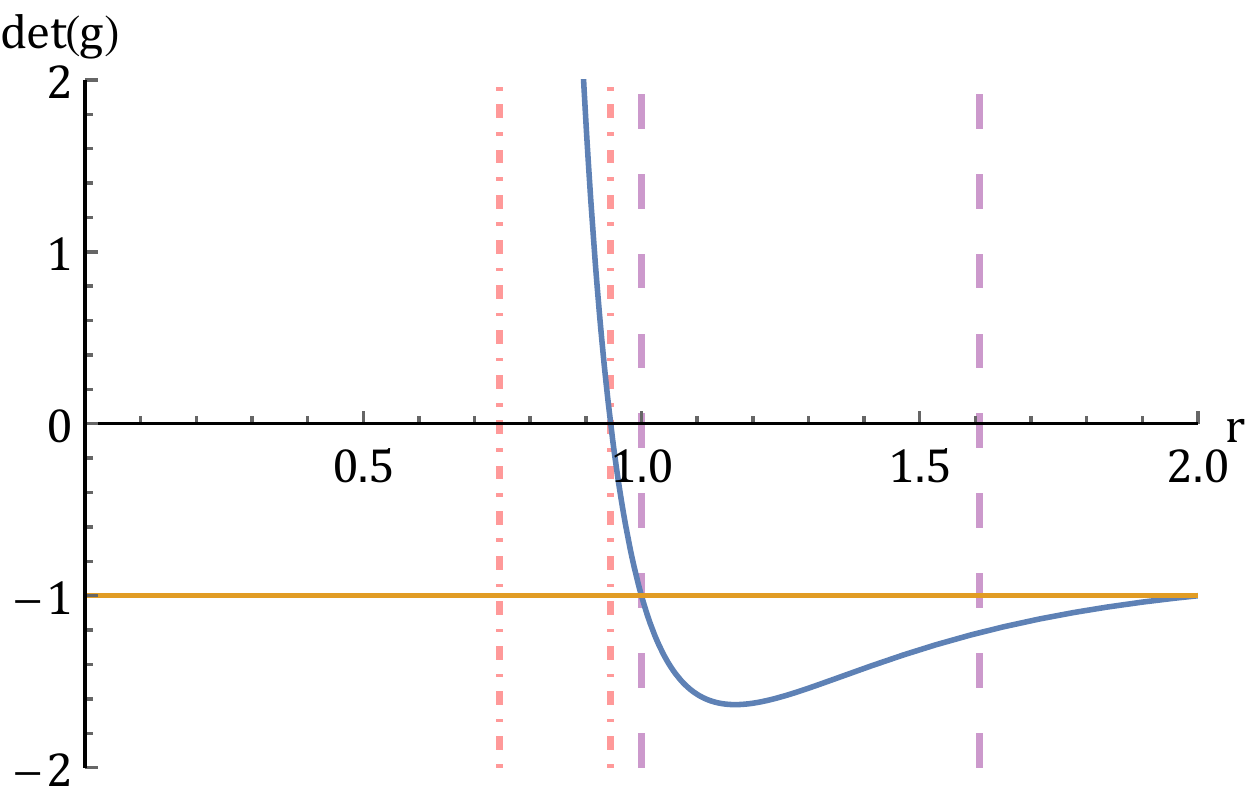}
 \includegraphics[scale=0.5]{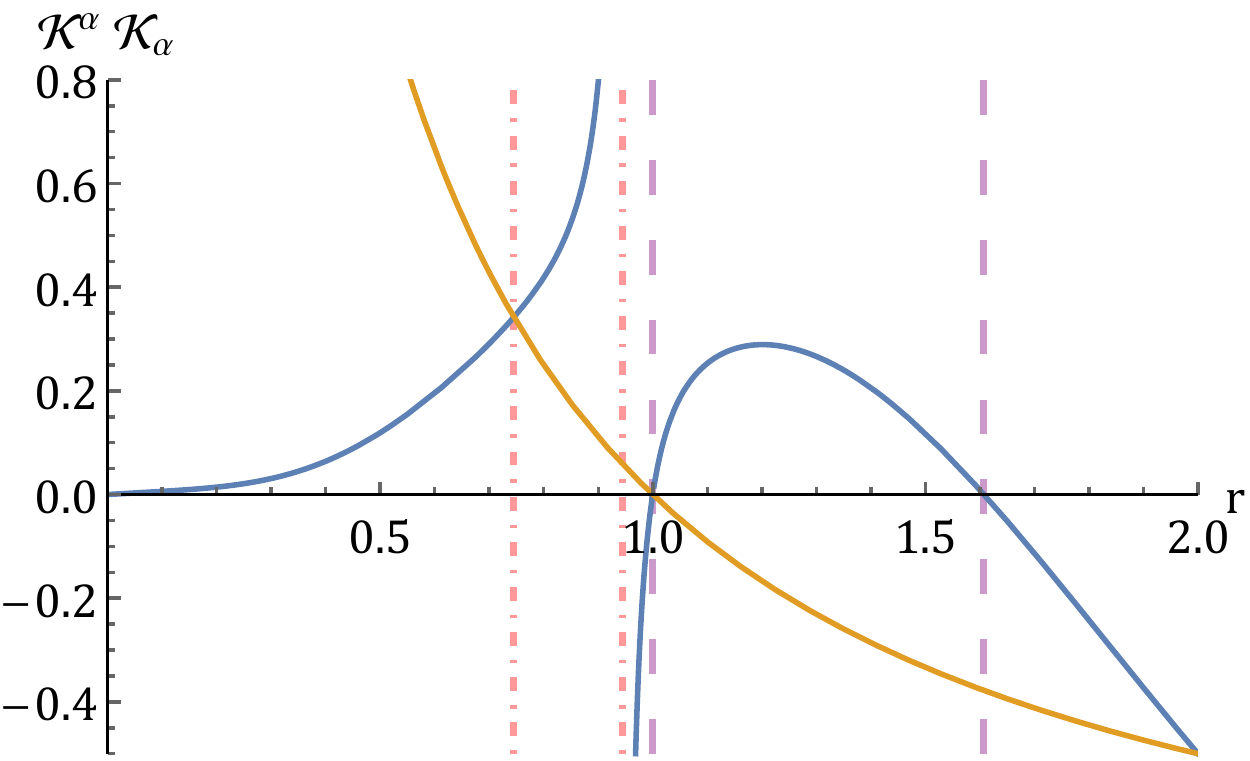}\,\,
 \includegraphics[scale=0.5]{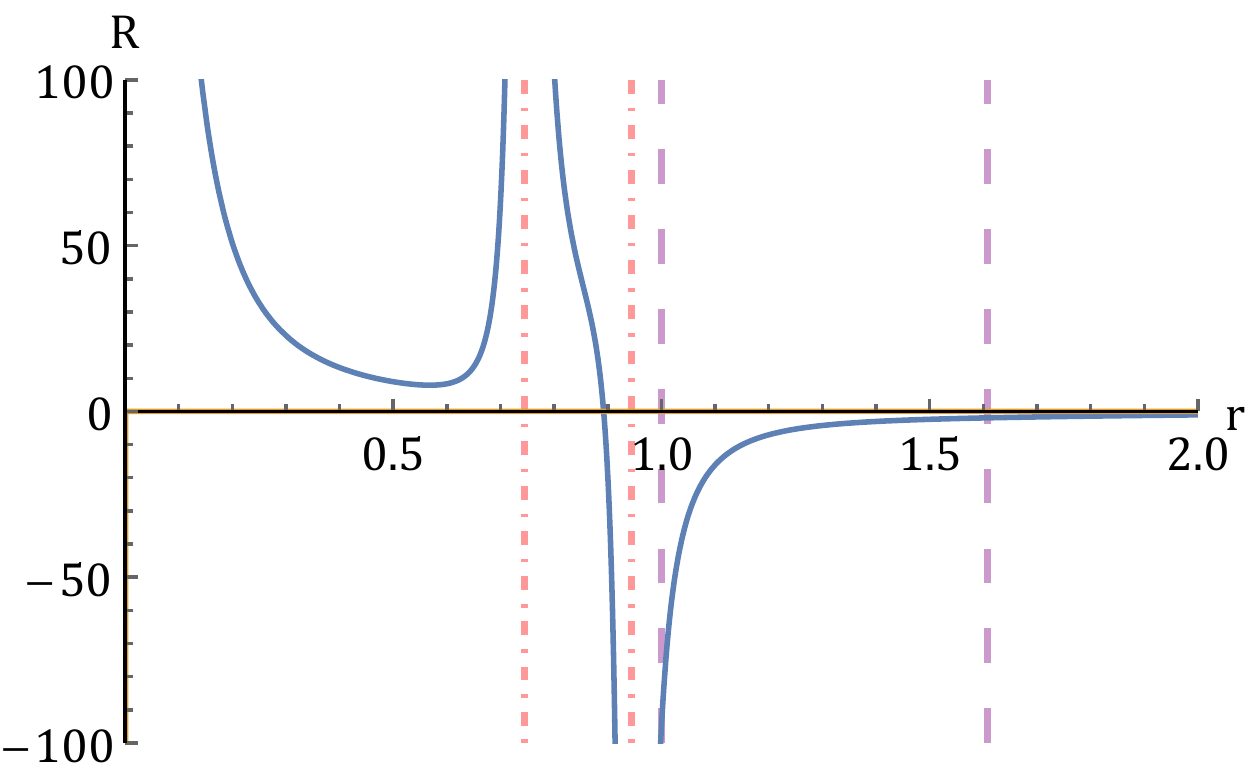}
  \caption{The determinant of the new metric $\text{det}(g)$, the $00$-components of the metric $g_{\alpha\beta}$,  the norm of the Kodama vector $\cK^{\alpha} \cK_{\alpha}$ as well as the Ricci scalar $R$ for the specific case $ \mu = 1$, $\eta/ \beta = 2$,  $\eta / (2\zeta) = 10^{-4}$ and $r_{\ast} /\mu= 4$. The yellow lines indicate the quantities related to the seed solution, while the blue lines indicate the behaviour of the modified geometry. The violet large dashed lines indicate the location of the two horizons, while the orange dashed lines indicate singularities.}
    \label{fig1}
\end{figure}

\begin{figure}[h!]
  \centering
  \includegraphics[scale=0.5]{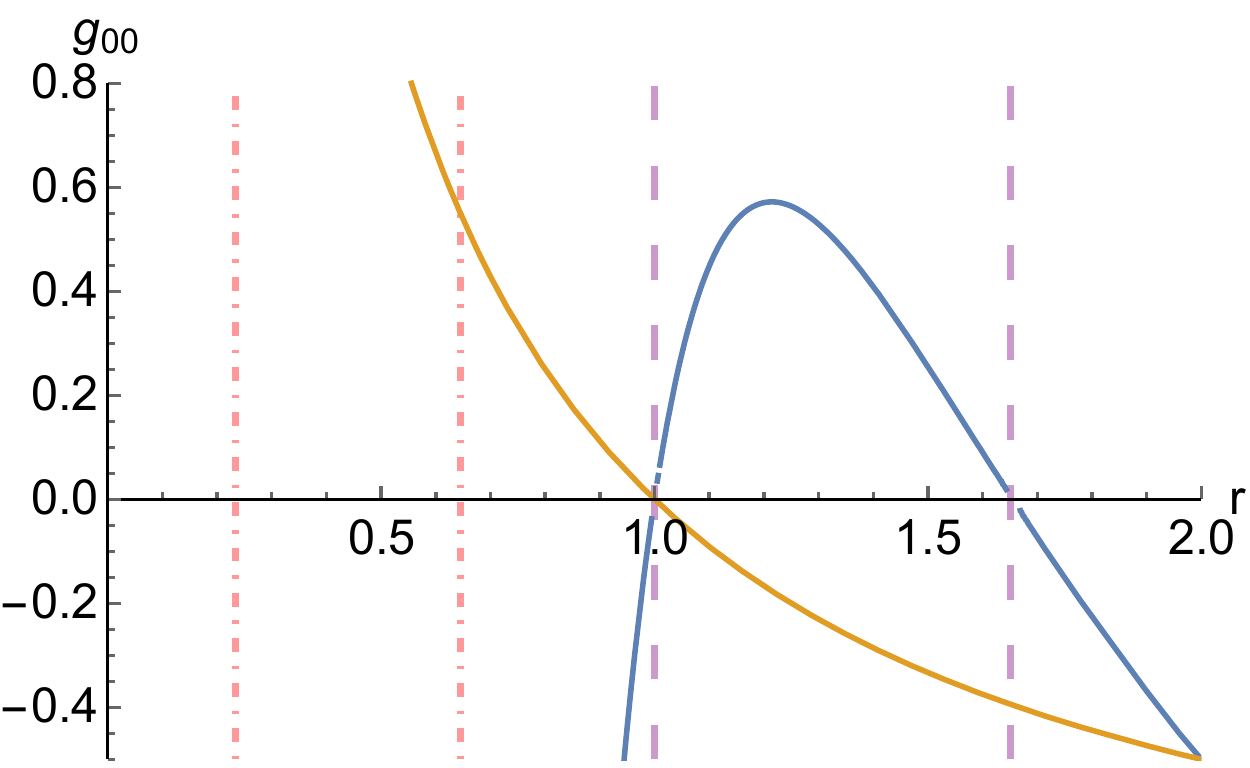}
 \; \includegraphics[scale=0.5]{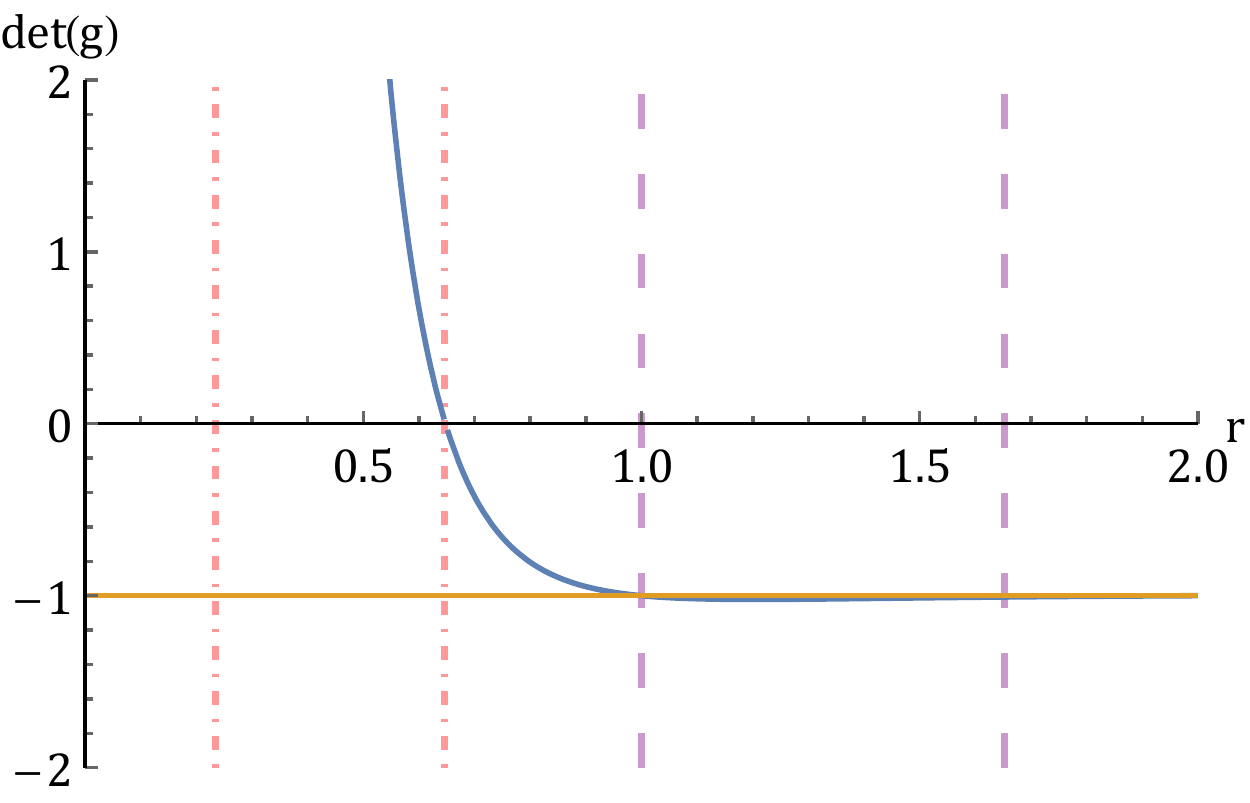}
 \includegraphics[scale=0.5]{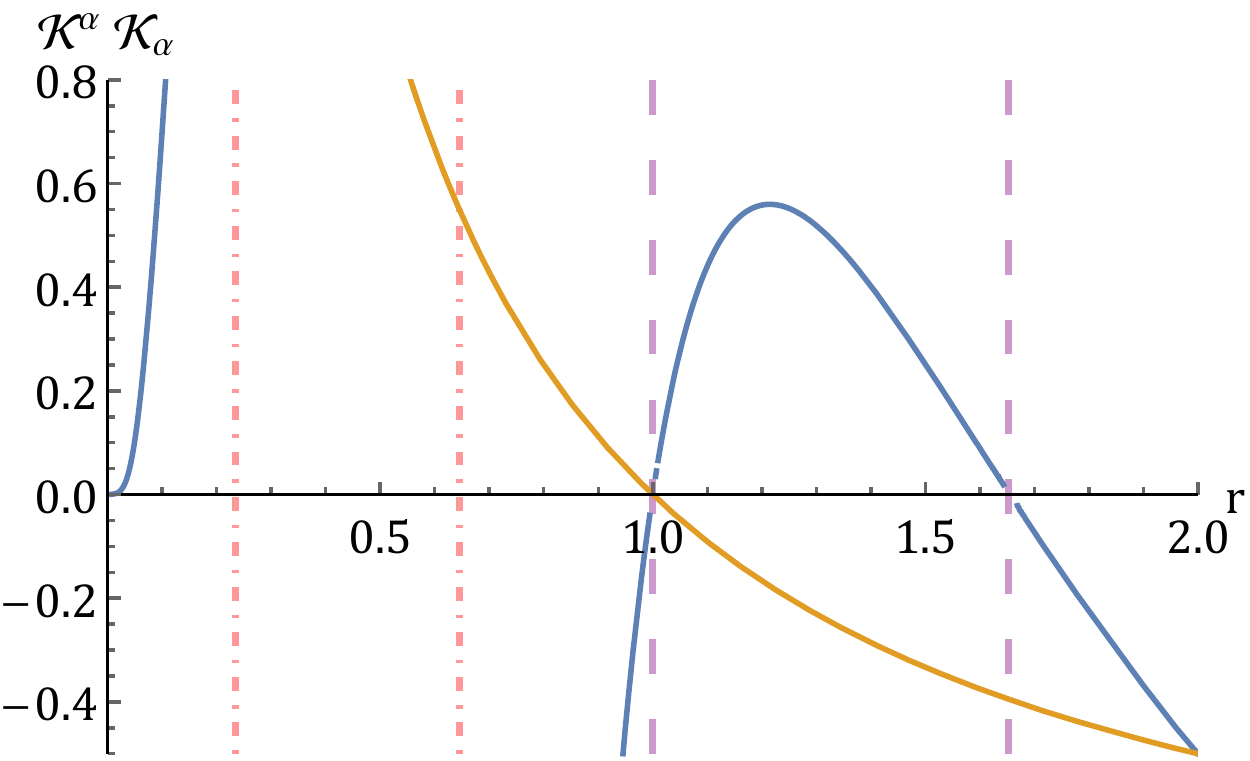}\,\,
 \includegraphics[scale=0.5]{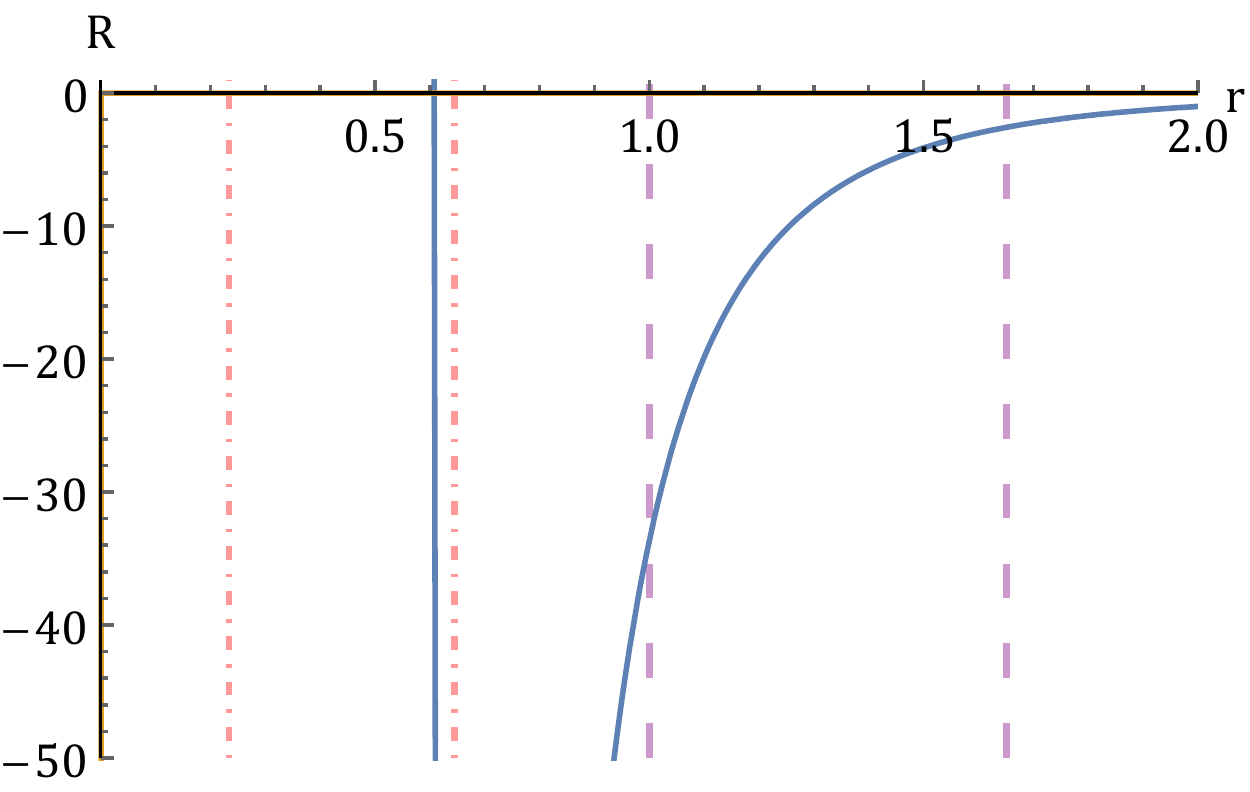}
  \caption{The determinant of the new metric $\text{det}(g)$, the $00$-components of the metric $g_{\alpha\beta}$,  the norm of the Kodama vector $\cK^{\alpha} \cK_{\alpha}$ as well as the Ricci scalar $R$ for the specific case $ \mu = 1$, $\eta/ \beta = 200$,  $\eta / (2\zeta) = 10^{-4}$ and $r_{\ast} /\mu= 4$. The yellow lines indicate the quantities related to the seed solution, while the blue lines indicate the behaviour of the modified geometry. The violet large dashed lines indicate the location of the two horizons, while the orange dashed lines indicate singularities.}
    \label{fig2}
\end{figure}

As a first example, we consider the case $\eta/ \beta =2$. 
The plots of the three key quantities are given in Figure-\ref{fig1}. First, one can observe that the norm of the Kodama vector vanishes only one time in the seed solution, while it has two zeros in the modified geometry, signaling thus two horizons. The plot of the $00$-component of the metric shows that it vanishes precisely at the locus of each horizons, showing that they are indeed light-like horizons. Moreover it is clear from the sign of the norm of the Kodama vector that it becomes space-like in two regions, which signals the existence of a trapped region, as expected. The Ricci scalar remains regular on each horizon, and diverges in the interior region, where there are two new singularities, represented by the orange vertical lines. Nevertheless, they affect only the interior region bounded by the inner horizon, such that the geometry is well defined only up to the first singularity. In this range of $r$, the determinant is negative and keeps the same sign in the trapped region, ensuring that one can safely cross each horizons. Moreover, as the determinant does not vanish on any horizon, it implies from (\ref{det4}) that the kinetic term of the scalar field remains also regular on these horizons, providing a well defined hairy configurations.
Notice that the position of the singularity can be shifted by tuning the value of the ration $\eta/\beta$. This is depicted on the Figure-\ref{fig2}, which corresponds to a ratio $\eta/\beta = 200$ while keeping the same values for the other parameters. For this second example, the singularity is pushed in the deep interior of the black hole. 

As a final remark, notice that one can use (\ref{fonc44}) to write the potential $G_2(r)$ as a function of the kinetic term $\tilde{X}$ of the seed solution. We obtain therefore
\be
G_2(r(\tilde{X})) = \left[ 1 -  r_{\ast} \sqrt{\frac{ \eta}{\sqrt{2}\beta} \frac{\tilde{X}^{3/2}}{1 + q^2\tilde{X}}}\right]^2\,.
\ee 

\bigskip

Let us now investigate the case with $q=0$. In this static limit, the above modifications cannot be introduced anymore without spoiling the behavior of the determinant. One encounters the same situation as when starting from the JNW seed solution, as both are characterized by a non-constant kinetic term but a static scalar profile.
To see this, let us set $A=1$ and investigate possible corrections through the function $G_2(r)$. Now the norm of the Kodama vector and the expression of the determinant of the metric are given by the simple expressions
\begin{align}
\cK_{\alpha} \cK^{\alpha} & = - \frac{ F(r)}{G_2(r)}\,,\\
\text{det}(g)&  = - r^4 \sin^2{\theta} \; G_2(r)\,.
\end{align}
Therefore, the function $G_2(r)$ is constrained to satisfy
\be
0 < G_2(r) < + \infty\,.
\ee 
Moreover, the asymptotic flatness requires that $G(r) \sim 1$ when $r\rightarrow + \infty$.  While these conditions leave a large freedom in the choice of $G_2(r)$ to modify the exterior geometry of the black hole, they cannot generate a new horizon, since it would generate some inconsistencies in the behavior of the determinant. This illustrates the importance, additionally to the existence of a non-constant kinetic term, of a time-dependent scalar profile for the scalar field to introduce interesting modification of the seed solution through a disformal transformation. 

Notice that for this sub-case, the kinetic term of the scalar field takes a fairly simple form, i.e  $\tilde{X} \propto r^{-4}$ which allows to re-write straightforwardly the disformal potential in term of $\tilde{X}$. 
Once the function $G_2(r)$ is chosen, one can use the profile of the kinetic term of the seed solution to write
\be
G_2(r) = G_2\left( \left(\frac{2\beta^2}{\eta^2\tilde{X}}\right)^{1/4}\right)\,,
\ee
which ensures that the transformation is invertible. This concludes our investigation of the disformal transformation of the last example of seed solutions.

%As already mentioned at the end of section~\ref{subsubsec:RNnonstealth}, we shall not require the theory after the disformal transformation to satisfy the phenomenological constraint summarized in section~\ref{subsec:consraints}. The black hole solution with $q \ne 0$ constructed in the present paper may still be observationally viable if it can be embedded in a cosmological solution with $\phi = const$. Starting with a cosmological solution with $\phi = const$, a scalar hair may develop during a black hole formation. The scalar field near the formed black hole may then become stationary after a while, but it should take a long time for the scalar hair to propagate to the cosmological distances. In this case the black hole with a $q\ne 0$ scalar hair may be consistently embedded in a cosmological background with $\phi = const$ and then we do not need to impose the phenomenological constraints summarized in section~\ref{subsec:consraints}. With the lack of our good understanding of black hole formation in DHOST theories, we have to admit that this is just an expectation or a conjecture. Nonetheless it is at least an intriguing possibility. Dedicated study of the formation process is needed in order to understand the nature of black holes in DHOST theories. 

Black holes have been one of the most important tools to study theories of gravity. Therefore, besides the possibility of being a phenomenologically viable candidate for astrophysical black holes in our universe, it is by itself theoretically important to explore the space of black hole solutions. The new black hole solutions found in this section serves as one of few examples of non-stealth black hole solutions in DHOST theories.

\section{Discussion}
\label{sec6}

Inspired by the standard solution generating methods, we have investigated how disformal field redefinitions can be used to generate new spherically symmetric exact solutions for quadratic DHOST theories. Throughout this work, we have restricted ourselves to invertible disformal transformation and shift symmetric disformal potentials, i.e $A:=A(X)$ and $B:=B(X)$. We have tested this solution generating method on three illustrative examples of seed solutions. Let us summarize the outcome of this procedure on each case.

In the first part of this work, we have considered a seed solution of the Einstein-Scalar system with a canonical scalar field. Due to the no hair theorem, this system only exhibits solutions describing naked singularity in the static spherically symmetric sector. We have therefore considered the simplest example of naked singularity solutions known as the Janis-Newman-Winicour (JNW) exact solution (\ref{JNW}). Under rather general assumptions, we have shown that the new family of exact solutions obtained from a disformal mapping of the JNW solution fails to contain black hole solutions.  This first result underlines the difficulty to generate a well defined black hole solution from a seed solution describing an horizonless compact object. In particular, while dressing the naked singularity with an apparent horizon is straightforward, such horizon fails to be light-like and regular and thus, to describe a black hole horizon. 

In the second part of this work, we have used the stability properties of the degeneracy classes of quadratic DHOST theories to generate new exact solutions starting from known black hole solutions of DHOST theories. We have considered two examples of exact solutions associated to a time-dependent scalar field profile as seed solutions. The first one consists of the simplest stealth Schwarzschild black hole associated to constant scalar kinetic term $\tilde{X} = \tilde{X_{\circ}}$. The strong coupling problem~\cite{deRham:2019gha} around the stealth solution can be removed by considering approximately stealth solutions, as already known in the context of ghost condensation~\cite{Mukohyama:2005rw,Mukohyama:2009rk,Mukohyama:2009um} and as will be soon reported also in DHOST theories~\cite{Motohashi:2019ymr}. This renders the stealth solution a viable seed solution. The second seed solution is a non-stealth black hole solution of shift symmetric GLPV theories presented in \cite{Babichev:2017guv} which exhibits a radially dependent scalar kinetic term $\tilde{X}(r)$.

Starting form the stealth Schwarzschild solution, we have shown that the property of having a constant kinetic term in the seed configuration severely restricts the modifications one can introduce through the disformal potentials. Indeed,
the disformally related solution is also associated to a constant kinetic term. As a result, the disformal potentials $A$ and $B$ are also forced to be constant in spacetime even if they are non-trivial functions of $X$. However, despite this restriction, one obtains two different types of solutions. When $\tilde{X}_{\circ} \neq -q^2$, one obtains a non stealth black hole which is asymptotically locally flat but which exhibits a deficit solid angle due to the non-vanishing spatial gradient of the scalar field at $r\rightarrow +\infty$. When $\tilde{X}_{\circ} = -q^2$, one obtains another stealth  Schwarzschild black hole, and the disformal mapping shifts only the value of its mass. This difference is made clear by the expression of the Misner-Sharp energy in each case.

Then, starting from the non-stealth solution associated to a radially dependent scalar kinetic term, we have obtained yet another family of exact solutions. The combined properties of the seed solution of having a time-dependent scalar profile as well as a non-constant kinetic term opens up as expected new ways to modify the geometry through radially dependent disformal potentials, and allows to generate new horizons on top of the one present in the seed solution. We point out that if one removes the time-dependence of the scalar profile, this is no longer true, underlining the importance of having both a non-constant kinetic term as well as the time-dependent scalar profile. Only few non-stealth solutions of DHOST with such configurations are known, and they could therefore provide interesting starting points to build new exact solutions of interest for DHOST theories. See \cite{Babichev:2013cya, Kobayashi:2014eva} for examples.

Another useful application of this generating solution-method is to consider a small transformation which allows to build minimally modified Schwarzschild black hole solution, starting from the stealth configuration. We have presented the outcome of such construction at the end of Section-\ref{secstealth}. This provides an interesting way to parameterize small deviations from the GR vacuum solution, which in turn, are exact solutions, at leading order in the disformal parameters, of some sub-sector of DHOST theories. An interesting application of this idea would be to construct minimally modified Kerr black hole solutions in DHOST theories and investigate their phenomenology.

Let us point that, due to the recent observational constraints imposing $c_{g}=c$, one needs to make sure that the new solutions presented in this work are indeed solutions of DHOST theories satisfying these observational constraints. It turns out for the DHOST theories admitting the stealth Schwarzschild solution, this is indeed possible, as the seed solution considered above in the context of DHOST theories solve the field's equation of theories which does not satisfy these observational constraints, but which can be mapped to the subsector satisfying $c_{g}=c$ after the disformal mapping. The conditions for such mapping have been discussed in Section-\ref{sec1.3}. However, in the case of the non-stealth seed solution, we supposed that the resulting solution is embedded in a cosmological solution with $\phi=const$ for time scales relevant for astrophysical observations and, for this reason, we did not need to impose the conditions discussed in Section-\ref{sec1.3}.

As a final word, we emphasize that this work represents only a first step towards investigating this solution generating method based on disformal transformations of known solutions. One can now use this solution generating method and explore the solution space of DHOST theories beyond the stealth sector in an efficient way. An interesting application would be to generate new rotating exact solutions beyond the stealth sector starting from the stealth Kerr solution presented in \cite{Charmousis:2019vnf}.

Finally, let us point that a crucial question is the relevance of these new exact solutions when considering matter coupling. Indeed, in order to observe the effect of these modified gravity theories on matter beyond the test field approximation, it is crucial to consider seed solutions which contain non-trivial matter. One example would be to consider one of the exact solutions of the Einstein-Maxwell-Dilaton system, which would allow to investigate the effect of the DHOST higher order term on the dynamics of the Maxwell field. We leave this question for future work. \bigskip

\newpage
%===============================================================================
\textbf{Acknowledgments}\smallskip

The authors would like to thank D.~Langlois, H.~Motohashi and K.~Noui for useful discussions. The work of JBA was supported by Japan Society for the Promotion of Science Grants-in-Aid for Scientific Research No. 17H02890. The work of SM was supported by Japan Society for the Promotion of Science Grants-in-Aid for Scientific Research No. 17H02890, No. 17H06359, and by World Premier International Research Center Initiative, MEXT, Japan. 
%===============================================================================

%%\bibliographystyle{bib-style}

\end{document}